\documentclass[sigconf,9pt,table]{acmart}
\usepackage[english]{babel}
\usepackage{blindtext}
\usepackage{xspace}
\usepackage{lipsum}
\usepackage{amsmath}
\usepackage{amsthm}

\usepackage{algorithm}
\usepackage{algpseudocode}

\usepackage[normalem]{ulem}
\usepackage{caption}
\usepackage{url}
\usepackage{float}
\usepackage{cases}
\usepackage{color}
\usepackage{comment}
\usepackage{subcaption}
\usepackage{array}
\usepackage{booktabs}
\usepackage{csquotes}
\usepackage[font=itshape]{quoting}
\usepackage{xcolor,colortbl}

\definecolor{Gray}{gray}{0.85}

\newcommand{\name}{{\textsc{Genet}}\xspace}

\newcommand{\jcstrike}[1]{}

\definecolor{dartmouth-green}{rgb}{0.05, 0.5, 0.06}

\newcommand{\zxedit}[1]{{\color{black}{#1\xspace}}}

\newcommand{\sid}[1]{{\color{cyan}{}}}

\newcommand{\sidstrike}[1]{}

\newcommand{\zx}[1]{{\color{black}{#1\xspace}}}

\newcommand{\disclaimer}{{Colored rows show the configurations (and their ranges) used in the simulator in the original paper.}\xspace}

\newcommand{\ignore}[1]{}

\newcounter{packednmbr}
\newenvironment{packedenumerate}{\begin{list}{\thepackednmbr.}{\usecounter{packednmbr}\setlength{\itemsep}{0.5pt}\addtolength{\labelwidth}{-4pt}\setlength{\leftmargin}{2ex}\setlength{\listparindent}{\parindent}\setlength{\parsep}{1pt}\setlength{\topsep}{3pt}}}{\end{list}}
\newenvironment{packeditemize}{\begin{list}{$\bullet$}{\setlength{\itemsep}{0.5pt}\addtolength{\labelwidth}{-4pt}\setlength{\leftmargin}{2ex}\setlength{\listparindent}{\parindent}\setlength{\parsep}{1pt}\setlength{\topsep}{3pt}}}{\end{list}}

\newcommand{\tightcaption}[1]{\vspace{-0.35cm}\caption{{\normalfont{\textit{{#1}}}}\vspace{-0.47cm}}}

\newcommand{\tightsection}[1]{\vspace{-0.1cm}\section{#1}\vspace{-0.1cm}}

\vspace{-0.2cm}

\newcommand{\eg}{{e.g.,}\xspace}
\newcommand{\ie}{{i.e.,}\xspace}

\newcommand{\myparashort}[1]{\vspace{0.1cm}\noindent{\bf {#1}}~}
\newcommand{\mypara}[1]{\vspace{0.1cm}\noindent{\bf {#1}:}~}
\newcommand{\myparaq}[1]{\vspace{0.1cm}\noindent{\bf {#1}?}~}

\newcolumntype{L}[1]{>{\raggedright\let\newline\\\arraybackslash\hspace{0pt}}m{#1}}
\newcolumntype{C}[1]{>{\centering\let\newline\\\arraybackslash\hspace{0pt}}m{#1}}
\newcolumntype{R}[1]{>{\raggedleft\let\newline\\\arraybackslash\hspace{0pt}}m{#1}}

\setcopyright{acmcopyright}

\acmYear{2022}\copyrightyear{2022}
\acmConference[SIGCOMM '22]{ACM SIGCOMM 2022 Conference}{August 22--26, 2022}{Amsterdam, Netherlands}
\acmBooktitle{ACM SIGCOMM 2022 Conference (SIGCOMM '22), August 22--26, 2022, Amsterdam, Netherlands}
\acmPrice{15.00}
\acmDOI{10.1145/3544216.3544243}
\acmISBN{978-1-4503-9420-8/22/08}

\algrenewcommand\algorithmicrequire{\textbf{Input:}}
\algrenewcommand\algorithmicensure{\textbf{Output:}}

\newcommand{\RewardFunc}{\ensuremath{R}\xspace}
\newcommand{\Policy}{\ensuremath{\pi}\xspace}

\newcommand{\Config}{\ensuremath{p}\xspace}

\begin{document}

\newcommand*{\affaddr}[1]{#1} 
\newcommand*{\affmark}[1][*]{\textsuperscript{#1}}


\newcommand{\authdag}{\textsuperscript{\textdagger}}
\newcommand{\authsec}{\textsuperscript{\textsection}}
\newcommand{\authp}{\textsuperscript{\textparagraph}}

\title{
  \fontsize{13pt}{15pt}\selectfont
  \name: Automatic Curriculum Generation for Learning Adaptation in Networking\\
}

\author{
{\Large
  Zhengxu Xia$^{1*}$,
  Yajie Zhou$^{2*}$,
  Francis Y. Yan$^3$,
  Junchen Jiang$^1$
}
}

\affiliation{
{\normalsize
  $^1$University of Chicago,
  $^2$Boston University,
  $^3$Microsoft Research
}
\vspace{8pt}
}

\renewcommand{\shortauthors}{Z. Xia, Y. Zhou, F. Y. Yan, J. Jiang}
\renewcommand{\shorttitle}{\name: Automatic Curriculum Generation for Learning
Adaptation...}


\begin{abstract}
As deep reinforcement learning (RL) showcases its strengths in networking, its pitfalls are also coming to the public's attention. Training on a wide range of network environments leads to suboptimal performance, whereas training on a narrow distribution of environments results in poor generalization.

This work presents \name, a new training framework for learning better RL-based network adaptation algorithms. \name is built on curriculum learning, which has proved effective against similar issues in other RL applications. At a high level, curriculum learning gradually feeds more ``difficult'' environments to the training rather than choosing them uniformly at random. However, applying curriculum learning in networking is nontrivial since the ``difficulty'' of a network environment is unknown. Our insight is to leverage traditional rule-based (non-RL) baselines: If the current RL model performs significantly worse in a network environment than the rule-based baselines, then further training it in this environment tends to bring substantial improvement. \name automatically searches for such environments and iteratively promotes them to training.  Three case studies---adaptive video streaming, congestion control, and load balancing---demonstrate that \name produces RL policies that outperform both regularly trained RL policies and traditional baselines.

\end{abstract}

\maketitle

\vspace{-6pt}

{\let \thefootnote\relax\footnote{{This work is accepted by SIGCOMM'22. \\ $^*$Both authors contributed equally to this research.}}}
\vspace{-18pt}


\vspace{8pt}
\tightsection{Introduction}

Many recent techniques based on deep reinforcement learning (RL) are now among the state of the arts for various networking and systems adaptation problems, including congestion control (CC)~\cite{aurora}, adaptive bitrate streaming (ABR)~\cite{pensieve}, load balancing (LB)~\cite{park}, wireless resource scheduling~\cite{chinchali2018cellular}, and cloud scheduling~\cite{decima}.
For a given distribution of training network environments (\eg network connections with certain bandwidth patterns, delay, and queue length), RL trains a policy to optimize performance over these environments.

However, these RL-based techniques face two challenges that can ultimately impede their wide use in practice: 
\begin{packeditemize}
\item {\bf Training in a wide range of environments:}
When the training distribution spans a wide variety of network environments (\eg a large range of possible bandwidth), an RL policy may perform poorly even if tested in the environments drawn from the same distribution as training.
\item {\bf Generalization:} RL policies trained on one distribution of synthetic or trace-driven environments may have poor performance and even erroneous behavior when tested in a new distribution of environments.
\end{packeditemize}
Our analysis in \S\ref{sec:motivate} will reveal that, across three RL use cases in networking, these challenges can cause well-trained RL policies to perform much worse than traditional rule-based schemes in a range of settings.

These problems are not unique to networking. 
\zx{In other domains (\eg robotics, gaming) where RL is widely used, it is also known that RL models have performance issues in both new environments drawn from the training distribution and new environments drawn from an unseen distribution~\cite{tobin17, kirk2021survey, narvekar2016source, zhu2020transfer, narvekar2020curriculum}}. There have been many efforts to address these issues by enhancing offline RL training or retraining a deployed RL policy online. 
Since updating a deployed model is not always possible or easy (\eg loading a new kernel module for congestion control or integrating an ABR logic into a video player), we focus on improving RL training offline.

A well-studied paradigm that underpins many recent techniques to improve RL training is {\em curriculum learning}~\cite{narvekar2020curriculum}.
Unlike traditional RL training that samples training environments in a random order, curriculum learning generates a training curriculum that gradually increases the difficulty level of training environments, resembling how humans are guided to comprehend more complex concepts. Curriculum learning has been shown to improve generalization~\cite{adr,adr2,paired} as well as {\em asymptotic} performance~\cite{weinshall2018curriculum,justesen2018illuminating}, namely the final performance of a model after training runs to convergence. Following an easy-to-difficult routine allows the RL model to make steady progress and reach good performance.

In this work, we present {\em \name, the first training framework that systematically introduces curriculum learning to RL-based networking algorithms}.
\name automatically generates training curricula for network adaptation policies.
The challenge of curriculum learning in networking is how to sequence network environments in an order that prioritizes highly {\em rewarding} environments where the current RL policy's reward can be considerably improved.  
Unfortunately, as we show in \S\ref{sec:curricula}, several seemingly natural heuristics to identify rewarding environments suffer from limitations.

\begin{packeditemize}
\item First, they use {\em intrinsic properties} of each environment (\eg shorter network or workload traces~\cite{decima} and smoother network conditions~\cite{robustifying} are supposedly easier), but these intrinsic properties fail to
indicate whether the current RL model can be improved in an environment.
\item Second, they use {\em handcrafted heuristics} which may not capture all aspects of an environment that affect RL training (\eg bandwidth smoothness does not capture the impact of router queue length on congestion control, or buffer length on adaptive video streaming).
Each new application (\eg load balancing) also requires a new heuristic.
\end{packeditemize}

The idea behind \name is simple: An environment is considered {\em rewarding} if the current RL model has a large {\em gap-to-baseline}, \ie how much the RL policy's performance falls behind a traditional {\em rule-based} baseline (\eg Cubic or BBR for congestion control, MPC or BBA for adaptive bitrate streaming) in the environment.
We show in \S\ref{subsec:design:curricula} that the gap-to-baseline of an environment is highly indicative of an RL model's potential improvement in the environment.
Intuitively, since the baseline already shows how to perform better in the environment, the RL model may learn to ``imitate'' the baseline's known rules while training in the same environment, bringing it on par with---if not better than---the baseline. 
On the flip side, if an environment has a small or even negative gap-to-baseline, chances are that the environment is intrinsically hard (a possible reason why the rule-based baseline performs badly), or the current RL policy already performs well and thus training on it is unlikely to improve performance by a large margin. 
\zx{A small gap-to-baseline might also arise when the rule-based baseline has poor performance yet the RL model still has a large room for improvement.
\name ignores this case, but we will discuss it in \S\ref{sec:discuss}.}

\begin{table*}[t]
\begin{centering}

\scalebox{0.9}{
\begin{tabular}{@{}L{3.5cm}L{6cm}L{4cm}L{4.53cm}@{}}
\toprule
{\bf Use case} & {\bf Observed state} (policy input) & {\bf Action} (policy output) & {\bf Reward} (performance) \\
\midrule
Adaptive Bitrate (ABR) Streaming & future chunk size, history throughput, current buffer length & bitrate selected for the \newline next video chunk & $\sum_i (\alpha\cdot \text{Rebuf}_i + \beta\cdot \text{Bitrate}_i $ \newline $+\ \gamma\cdot \text{BitrateChange}_i) / n$ \\ \midrule
Congestion Control (CC) & RTT inflation, sending/receiving rate, \newline avg RTT in a time window, min RTT & change of sending rate in \newline the next time window & $\sum_i (a\cdot \text{Throughput}_i$ \newline $+\ b\cdot \text{Latency}_i + c\cdot \text{LossRate}_i) / n$  \\ \midrule
Load Balancing (LB) & past throughput, current request size, number of queued requests per server & server selection for \newline the current request & $-\sum_i \text{Delay}_i / n$ \\
\bottomrule
\end{tabular}
}
\end{centering}
\vspace{13pt}
\tightcaption{RL use cases in networked systems. Default reward parameters:
$\alpha=-10$ (rebuffering in seconds),
$\beta=1$ (bitrate in Mbps),
$\gamma=-1$ (bitrate change in Mbps),
$a=120$ (throughput in kbps),
$b=-1000$ (latency in seconds),
$c=-2000$. Details in \ref{app:metrics}.}
\label{tab:examples}
\vspace{-2pt}
\end{table*}

\begin{figure}[t]
    \centering
    \includegraphics[width=0.92\columnwidth]{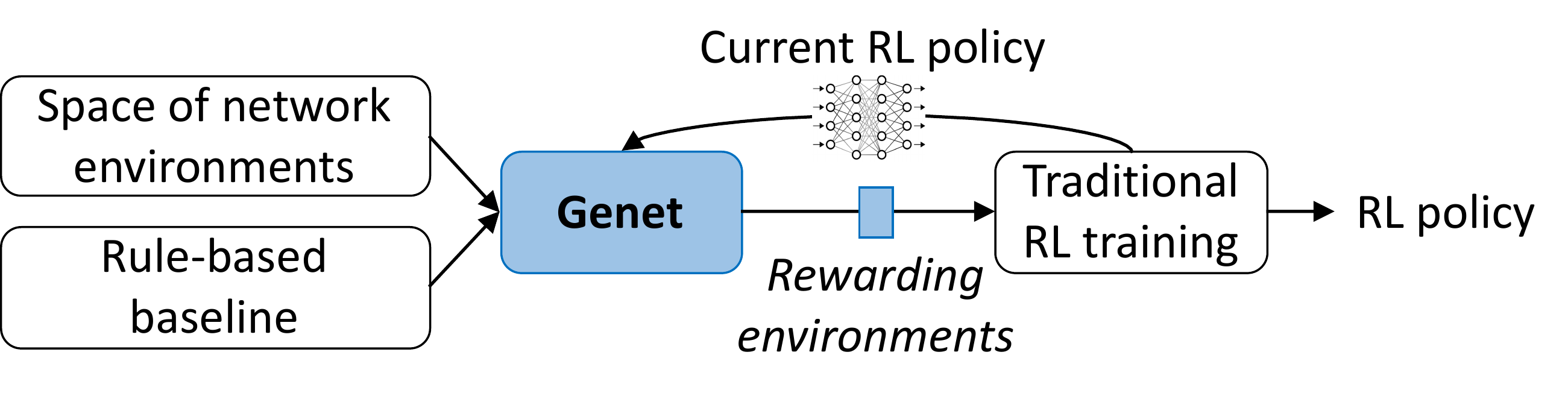}
    \tightcaption{\name creates training curricula by iteratively finding rewarding environments where the current RL policy has a large gap-to-baseline.}
    \label{fig:block-diagram}
    \vspace{-2pt}
\end{figure}

Inspired by the insight, \name generates RL training curricula by
iteratively identifying rewarding environments where the current RL model has a large gap-to-baseline 
and then adding them to RL training (Figure~\ref{fig:block-diagram}).
For each RL use case, \name parameterizes the network environment space, allowing us to search for rewarding environments in both synthetically instantiated environments and trace-driven environments. 
\name also uses Bayesian Optimization (BO) to facilitate the search in a large space. 
\zx{In particular, we cast the search for environments with a large gap-to-baseline as a maximum-search problem of a blackbox function in a high-dimensional space where each point represents a set of environment configurations and the function value is the gap-to-baseline. 
BO is then used to find a set of training environments with large gap-to-baselines.
}
\name is generic, since it does not use handcrafted heuristics to measure the difficulty of a network environment; instead, it uses rule-based algorithms, which are abundant in the literature of many networking and system problems, to generate training curricula.
Moreover, by focusing training on places where RL falls behind rule-based baselines, \name directly minimizes the chance of performance regressions relative to the baselines.
This is important, because system operators are more willing to deploy an RL policy if it outperforms the incumbent rule-based algorithm in production without noticeable performance regressions.\footnote{An example of this mindset is that a new algorithm must compete with the incumbent algorithm in A/B testing before being rolled out to production.} 

We have implemented \name as a separate module with a unifying abstraction that interacts with the existing codebases of RL training to iteratively select rewarding environments and promote them in the course of training. 
We have integrated \name with three existing deep RL codebases in the networking area---adaptive video streaming (ABR)~\cite{pensieve-code}, congestion control (CC)~\cite{aurora-code}, and load balancing (LB)~\cite{park-code}.

It stands to reason that \name is not without limitations. 
For instance, \name-trained RL policies might not outperform all rule-based baselines (\S\ref{subsec:eval:micro} shows that when using a naive baseline to guide \name, the resulting RL policy could still be inferior to stronger baselines).
\name-trained RL policies may also achieve undesirable performance in environments beyond the training ranges (\eg if we train a congestion-control algorithm on links with bandwidth between 0 and 100 Mbps, \name will not optimize for the bandwidth of 1 Gbps).
Moreover, \name does not guarantee adversarial robustness 
which sometimes conflicts with the goal of generalization~\cite{raghunathan2019adversarial}.

Using a combination of trace-driven simulation and real-world tests across three use cases (ABR, CC, LB), we show that \name improves asymptotic performance by 8--25\% for ABR, 14--24\% for CC, 15\% for LB, compared with traditional RL training methods. 
\zx{\name aims to optimize an RL model's asymptotic performance (\ie in-distribution generalizability), and it does not explicitly optimize the generalization in arbitrary test environments (\ie out-of-distribution generalizability). 
That said, our empirical test results show that \name-trained models improve not only asymptotic performance, but also the performance in unseen network environments. 
} 

\zx{The traces and scripts used in \name are released at {\url{https://github.com/GenetProject/Genet}}.}


\section{Motivation}
\label{sec:motivate}

Deep reinforcement learning (RL) trains a deep neural net (DNN) as the decision-making logic (policy) and is well-suited to many sequential decision-making problems in networking~\cite{park,haj2019view}.\footnote{There are rule-based alternatives to DNN-based policies, but they are not as expressive and flexible as DNNs, which limits their performance. Oboe~\cite{oboe}, for instance, sets optimal hyperparameters for RobustMPC based on the mean and variance of network bandwidth and as shown in \S\ref{subsec:eval:rule}, is a very competitive baseline, but it performs worse than the best RL strategy.}
We use three use cases (summarized in Table~\ref{tab:examples}) to make our discussion concrete: 
\begin{packeditemize}
\item An {\bf adaptive bitrate (ABR)} algorithm adapts the chunk-level video bitrate to the dynamics of throughput and playback buffer (input state) over the course of a video session. 
ABR policies, including RL-based ones (Pensieve~\cite{pensieve}), choose the next chunk's bitrate (output decision) at the chunk boundary to maximize session-wide average bitrate, while minimizing rebuffering and bitrate fluctuation. 
\item A {\bf congestion control (CC)} algorithm at the transport layer adapts the sending rate based on the sender's observations of the network conditions on a path (input state).
An example of RL-based CC policy (Aurora~\cite{aurora}) makes sending rate decisions at the beginning of each interval (of length proportional to RTT), to maximize the reward (a combination of throughput, latency, and packet loss rate). 
\item A {\bf load balancing (LB)} algorithm in a key-replicated distributed database reroutes each request to one of the servers (whose real-time resource utilization is unknown), based on the request arrival intervals, resource demand of past requests, and the number of outstanding requests currently assigned to each server.
\end{packeditemize}

We choose these use cases because they have open-source implementations (Pensieve~\cite{pensieve-code} for ABR, Aurora~\cite{aurora-code} for CC, and Park~\cite{park-code} for LB). Our goal is to improve existing RL training in networking. Revising the RL algorithm {\em per se} (input, output, or DNN model) is beyond our scope.

\mypara{Network environments}
We generate simulated training environments with a range of parameters, following prior work~\cite{pensieve,aurora,park}. 
An environment can be synthetically generated using a list of parameters as {\em configuration},
\eg in the context of ABR, a configuration encompasses bandwidth range, frequency of bandwidth change, chunk length, etc.
Meanwhile, when recorded bandwidth traces are available (for CC and ABR experiments), we can also create {\em trace-driven} environments where the recorded bandwidth is replayed. Note that bandwidth is only one dimension of an environment and must be complemented with other synthetic parameters in order to create a simulated environment.
(Our environment generator and a full list of parameters are documented in \S\ref{app:trace}.)
In recent papers, both trace-driven (\eg~\cite{pensieve,robustifying}) and synthetic environments (\eg~\cite{aurora,park}) are used to train RL-based network algorithms.
We will explain in \S\ref{subsec:design:framework} how our technique applies to both types of environments.

\mypara{Traditional RL training}
Given a user-specified distribution of (trace-driven or synthetic) training environments, the traditional RL training method works in iterations. 
Each iteration randomly samples a subset of environments from the provided distribution and then updates the DNN-based RL policy (via forward and backward passes). 
For instance, Aurora~\cite{aurora} uses an iteration of 7200 steps (\ie 30--50 30-second network environments) and applies the PPO algorithm to update the policy network by simulating the network environments in each batch.

Several previous efforts have demonstrated the promise of the traditional RL training---given the distribution of target environments, an RL policy can be trained to perform well in these environments (\eg~\cite{pensieve,aurora}). 
Unfortunately, this approach falls short on two fronts.

\begin{figure*}[t!]
    \begin{subfigure}[b]{0.48\linewidth}
    {
        \centering
        \includegraphics[width=1\linewidth]{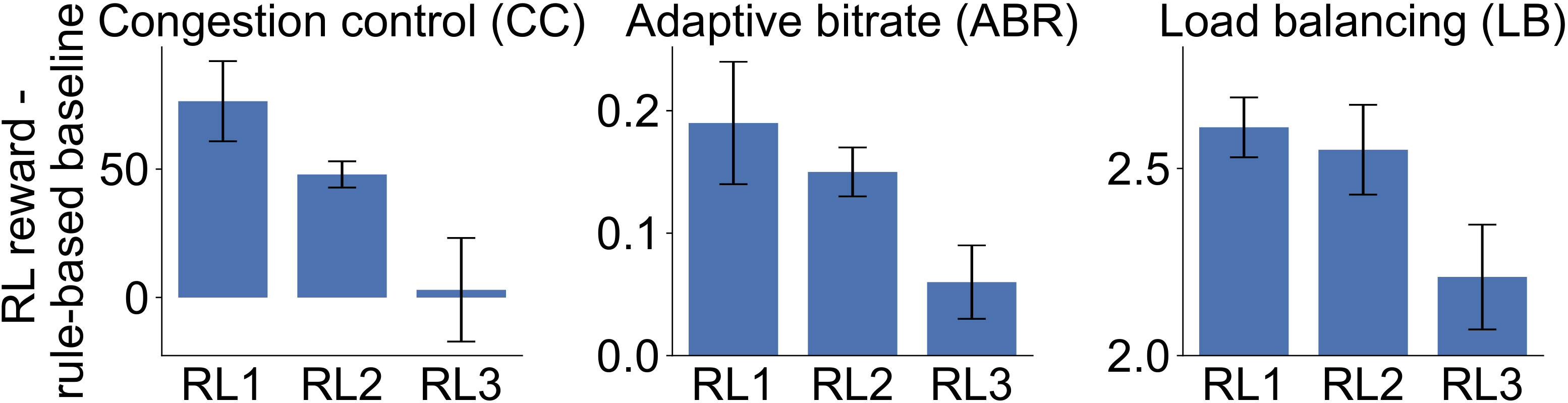}
        \caption{Performance gains of RL schemes over the baselines diminish as the target distribution spans a wide range of environments.}
        \label{}
    }
    \end{subfigure}
    \hfill
    \begin{subfigure}[b]{0.48\linewidth}
    {
        \centering
        \includegraphics[width=1\linewidth]{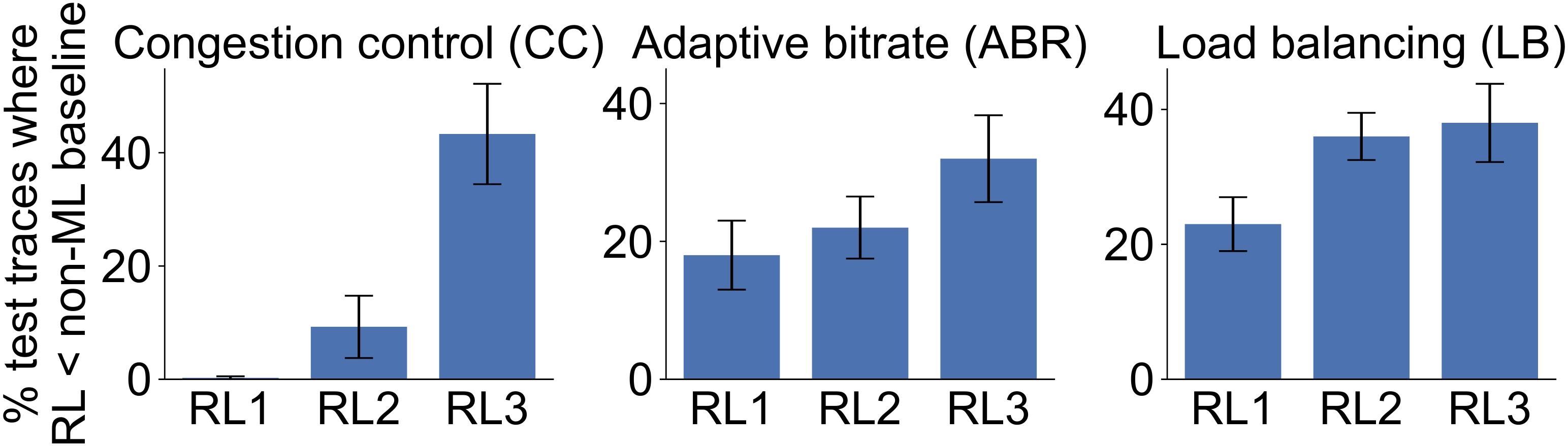}
        \caption{Even if RL schemes perform better on average, they are worse than the baselines on a substantial fraction of test environments.}
        \label{}
    }
    \end{subfigure}
    \vspace{3pt}
    \tightcaption{Challenges of RL training over a wider range of environments from small (RL1), medium (RL2), to large (RL3).}
    \vspace{10pt}
    \label{fig:asymptotic}
\end{figure*}

\begin{figure*}[t!]
    \centering
    \begin{subfigure}[b]{0.48\textwidth}
    {
        \centering
        \includegraphics[height=70pt]{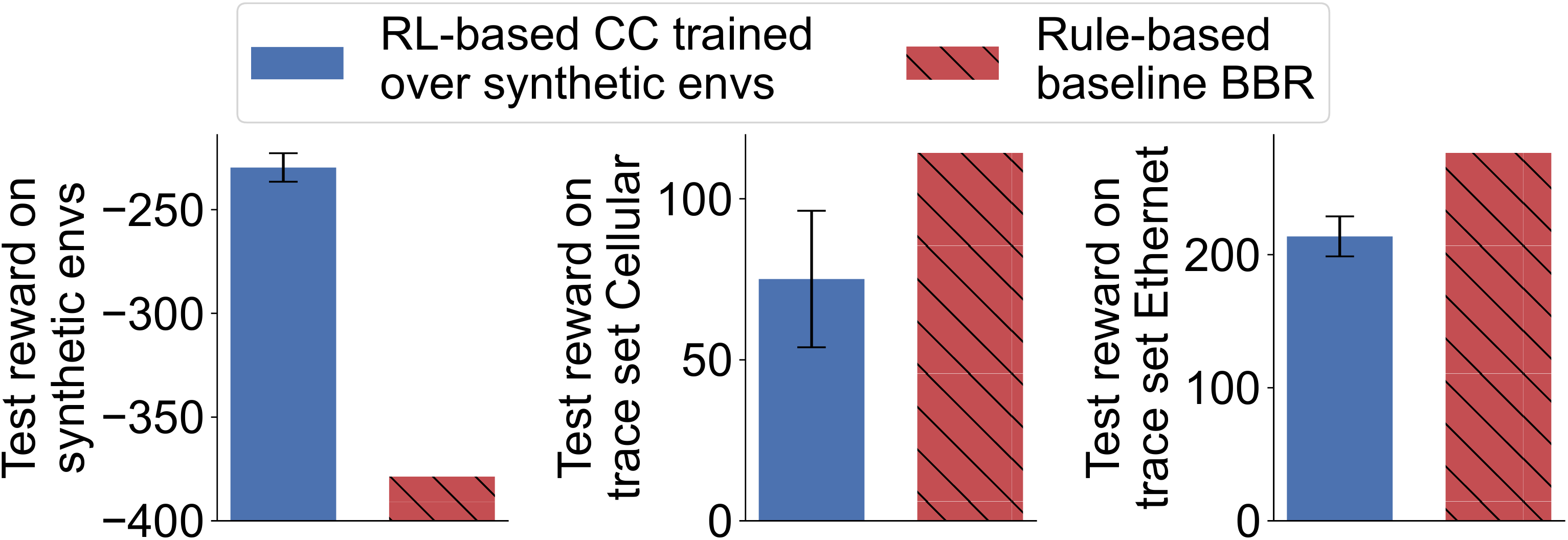}
        \caption{RL-based CC trained in synthetic network environments performs worse on real network traces than the rule-based baseline.}
        \label{}
    }
    \end{subfigure}
    \hfill
    \begin{subfigure}[b]{0.48\textwidth}
    {
        \centering
        \includegraphics[height=70pt]{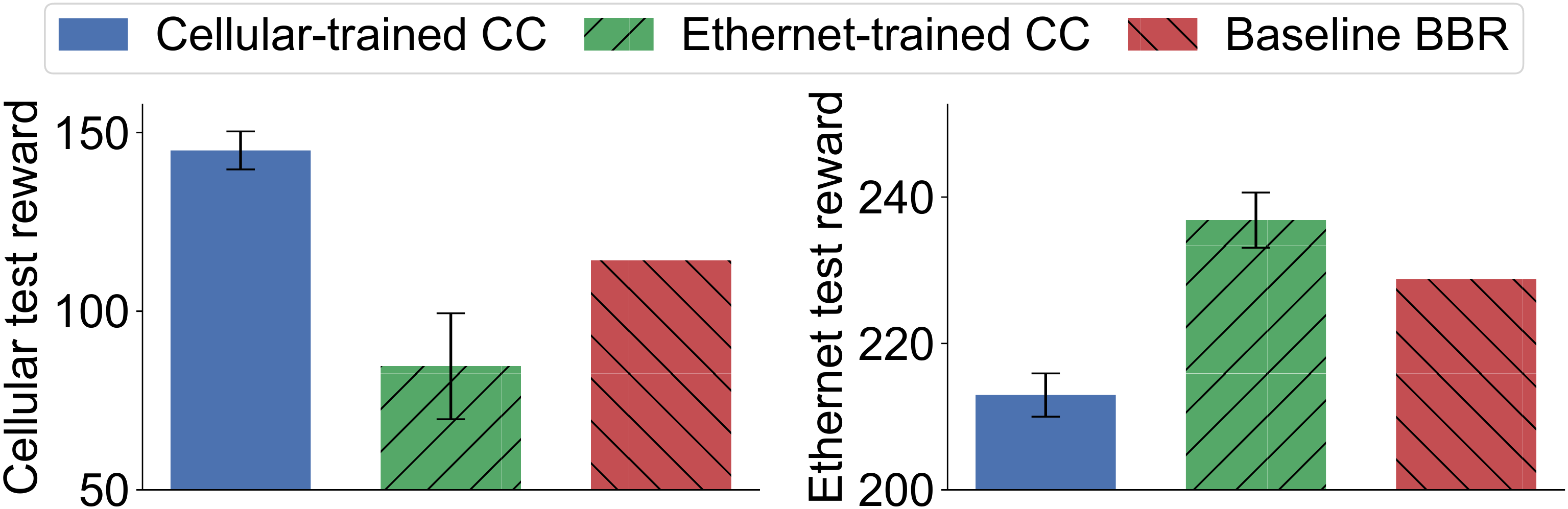}
        \caption{RL-based CC trained over one real trace set performs worse on another real trace set than the rule-based baseline.}
        \label{}
    }
    \end{subfigure}
    \vspace{3pt}
    \tightcaption{Generalization issues of RL-based schemes using CC as an example.}
    \vspace{5pt}
    \label{fig:generalize}
\end{figure*}

\myparashort{Challenge 1: Training over wide environment distributions.} \\
When the training distribution of network environments has a widespread (\eg a large range of possible bandwidth values), RL training tends to result in poor {\em asymptotic} performance (model performance after reaching convergence) even when the test environments are drawn from the {\em same} distribution as training.

In Figure~\ref{fig:asymptotic}, for each use case, we choose three target distributions (with increasing parameter ranges), labeled RL1/RL2/RL3 ranges of synthetic environment parameters in Table~\ref{tab:ABR-params}, \ref{tab:CC-params}, and \ref{tab:LB-params}.
Figure~\ref{fig:asymptotic}(a) compares the asymptotic performance of three RL policies (with different random seeds) with rule-based baselines, MPC~\cite{mpc} for ABR, BBR~\cite{bbr} for CC, and least-load-first (LLF) policy for LB, in test environments randomly sampled from the {\em same} ranges. It shows that RL's performance advantage over the baselines diminishes rapidly when the range of target environments expands. Even though RL-based policies still outperform the baselines on average, Figure~\ref{fig:asymptotic}(b) reveals a more striking reality---their performance falls behind the baselines in a substantial fraction of test environments. 

An intuitive explanation is that in each RL training iteration, only a batch of randomly sampled environments (typically 20--50) is used to update the model, and when the entire training set spans a wide range of environments, the batches between two iterations may have dramatically different distributions which potentially push the RL model to different directions.
This causes the training to converge slowly and makes it difficult to obtain a good policy~\cite{narvekar2020curriculum}. 
\zx{Although this problem is not completely avoided in our solution, it is mitigated by curriculum learning which draws the environments of a batch from a ``narrower'' training environment distribution, thus reducing the discrepancies between batches.}

\myparashort{Challenge 2: Low generalizability.}
Another practical challenge arises when the training process does not have access to the target environment distribution. 
This calls for models with good generalization, \ie the RL policies trained on one distribution also perform well on a different environment distribution during testing.
Unfortunately, existing RL training methods often fall short of this ideal.
Figure~\ref{fig:generalize} evaluates the generalizability of RL-based CC schemes in two ways.
\begin{packeditemize}
\item First, we train an RL-based CC algorithm on the same range of synthetic environments as specified in its original paper~\cite{aurora}. 
We first validate the model by confirming its performance against a rule-based baseline BBR, in environments that are independently generated from the same range as training (Figure~\ref{fig:generalize}(a); left).
Nevertheless, when tested on real-world recorded network traces under the category of ``Cellular'' and ``Ethernet'' from Pantheon~\cite{pantheon} (Table~\ref{tab:traces}), the RL-based policy yields much worse performance than the rule-based baseline.
\item Second, we train the RL-based CC algorithm on the ``Cellular'' trace set and test it on the ``Ethernet'' trace set (Figure~\ref{fig:generalize}(b); left), or vice versa (Figure~\ref{fig:generalize}(b); right).
Similarly, its performance degrades significantly when tested on a different trace set.
\end{packeditemize}
The observations in Figure~\ref{fig:generalize} are not unique to CC. 
Prior work~\cite{robustifying} also shows a lack of generalization of RL-based ABR algorithms.

\mypara{Summary}
In short, we observe two challenges faced by the traditional RL training mechanism:
\begin{packeditemize}
\item The asymptotic performance of the learned policies can be suboptimal, especially when they are trained over a wide range of environments. 
\item The trained RL policies may generalize poorly to unseen network environments. 
\end{packeditemize}


\tightsection{Curriculum learning for networking}
\label{sec:curricula}

Given these observations regarding the limitations of RL training in networking, a natural question to ask is {\em how to improve RL training such that the learned adaptation policies achieve good asymptotic performance across a broad range of target network environments.}\footnote{An alternative is to retrain the deployed RL policy whenever it meets a new domain (\eg a new network connection with unseen characteristics), but this does not apply when the RL policy cannot be updated frequently. Besides, it is also challenging to precisely detect model drift in the network conditions that necessitate retraining the RL policy.}

\mypara{Curriculum learning}
We cast the training of RL-based network adaptation to the well-studied framework of curriculum learning. Unlike the traditional RL training that samples training environments from a {\em fixed} distribution in each iteration, curriculum learning {\em varies} the training environment distribution to gradually increase the difficulty of training environments, so that training will see more environments {\em that are more likely to improve}, which we refer to as {\em rewarding} environments.
In many RL applications, prior work has shown the promise of curriculum learning, including faster convergence, higher asymptotic performance, and better generalization (\S\ref{sec:related}). 

\zx{The theoretical intuition behind curriculum learning is that a curriculum allows the model to optimize a family of gradually less smooth loss functions and prevents it from being trapped in local minima~\cite{bengio2009curriculum}. In the early stage of the curriculum, easier training samples are selected to comprise a smoothed loss function that reveals the big picture and is easier to optimize. The resulting model serves as a good starting point when more difficult samples are introduced to the training, reducing the smoothness of the loss function and making it harder to optimize.
By optimizing the model on a sequence of loss functions with decreasing smoothness, the curriculum is able to gradually bring the model parameters close to the global optimum.}

However, the challenge of employing curriculum learning lies in determining which environments are rewarding. Apparently, the answer to this question varies with applications, but three general approaches exist: 
(1) training the current model on a set of environments individually to determine in which environment the training progresses faster;
(2) using heuristics to quantify the easiness of achieving model improvement an environment; and 
(3) jointly training another model (typically DNN) to select rewarding environments. 
Among them, the first option is prohibitively expensive and thus not widely used, whereas the third introduces the extra complexity of training a second DNN. Therefore, we take a pragmatic stance and explore the second approach, while leaving the other two for future work. 

\mypara{Why sequencing training environments is difficult}
\zx{A common strategy in curriculum learning for RL is to measure environment difficulty and gradually introduce more difficult environments to training. To motivate our design choices, we first introduce three strawman approaches, with different strengths and weaknesses. 
They are used to determine how rewarding an environment is. 
A good approach should always select network environments in which the RL model has a large improvement in reward when trained in them.}

{\em Strawman 1: inherent properties.} The first idea is to quantify the difficulty level of an environment using some of its inherent properties. 
In congestion control, for instance, network traces with higher bandwidth variance are intuitively more difficult. 
This approach, however, only distinguishes environments that differ in the hand-picked properties and may not suffice under complex environments (\eg adding bandwidth traces with similar variance to training can have different effects).

{\em Strawman 2: performance of rule-based baselines.} 
Alternatively, one can use the test performance of a traditional algorithm to indicate the difficulty of an environment. Lower performance may suggest a more difficult environment~\cite{weinshall2018curriculum}. 
While this method can distinguish any two environments, it does not hint at how to improve the {\em current} RL model during training.

\zxedit{{\em Strawman 3: performance gap to the optimum.}
To fix the problem of Strawman 2, one can use the performance gap between the current RL policy and the optimum instead~\cite{robustifying}.
If the current model performs much worse than the optimum in an environment (\eg obtained by using ground-truth bandwidth as the bandwidth prediction), its performance might improve when trained in this environment.
A caveat of this approach is that the computation of the optimal performance could be prohibitively expensive or even infeasible. This approach may also fail to improve RL's performance in environments that are inherently hard (\eg highly fluctuating bandwidth in ABR and CC). 
}

\begin{figure}[t!]
    \begin{subfigure}[t]{0.49\linewidth}
    {
        \includegraphics[height=0.9\linewidth]{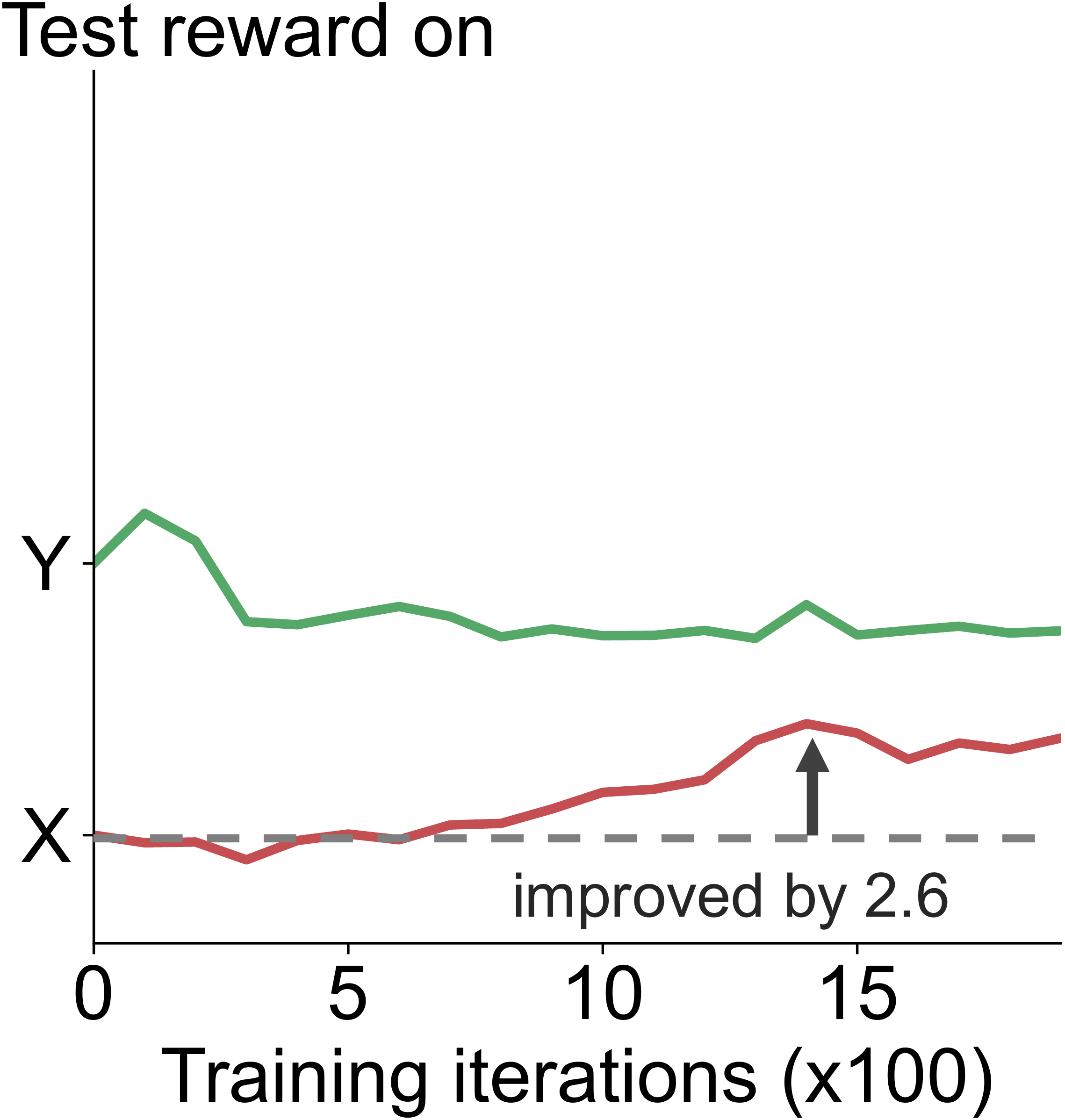}
        \label{fig:test-on-normal}
        \caption{Add X to training}
    }
    \end{subfigure}
    \hfill
    \begin{subfigure}[t]{0.49\linewidth}
    {
        \includegraphics[height=0.9\linewidth]{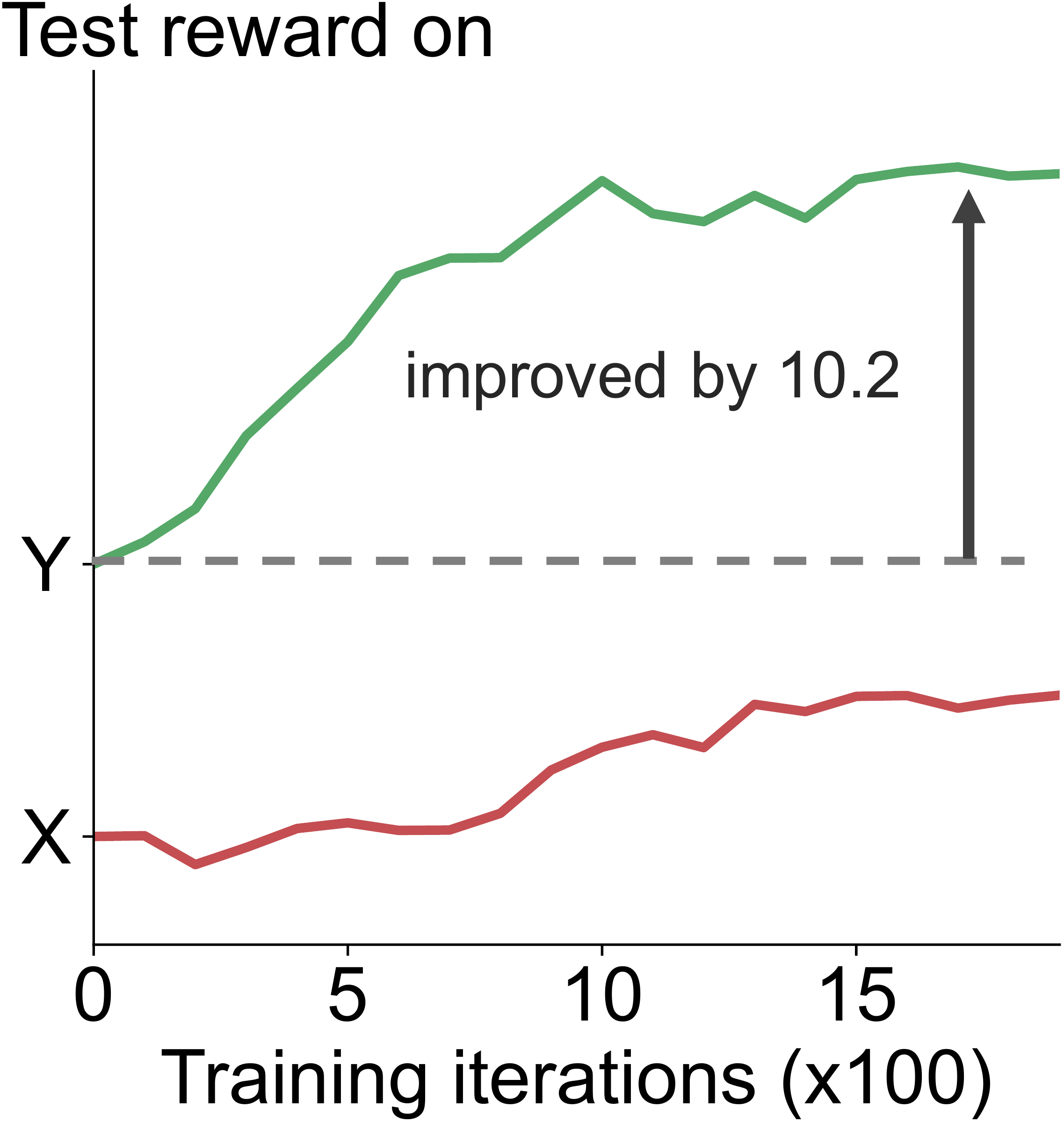}
        \label{fig:test-on-hard}
        \caption{Add Y to training}
    }
    \end{subfigure}
    \vspace{2pt}
    \tightcaption{A simple example where adding trace set X to training has a different effect than adding Y.
    Adding X to training improves performance on X only marginally but hurts Y, whereas adding Y improves the performance on both X and Y.
    }
    
    \label{fig:motivation-optimal}
\end{figure}
\vspace{-1pt}
\begin{figure}[t!]
    \centering
    \begin{subfigure}[t]{0.49\linewidth}
    {
        \includegraphics[height=0.9\linewidth]{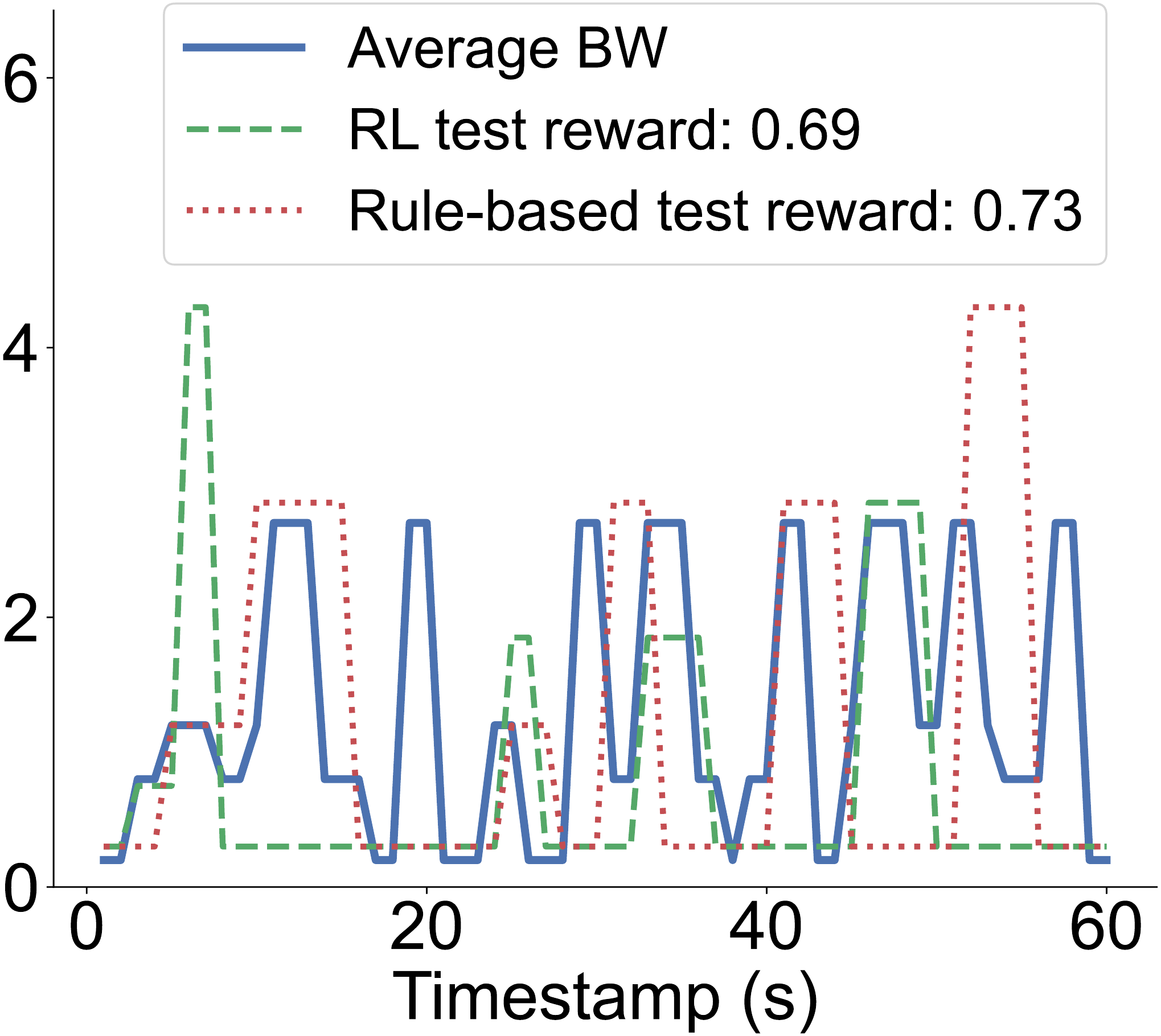}
        \caption{Trace in X (hard)}
        \label{fig:trace-example-y}
    }
    \end{subfigure}
    \begin{subfigure}[t]{0.49\linewidth}
    {
        \includegraphics[height=0.9\linewidth]{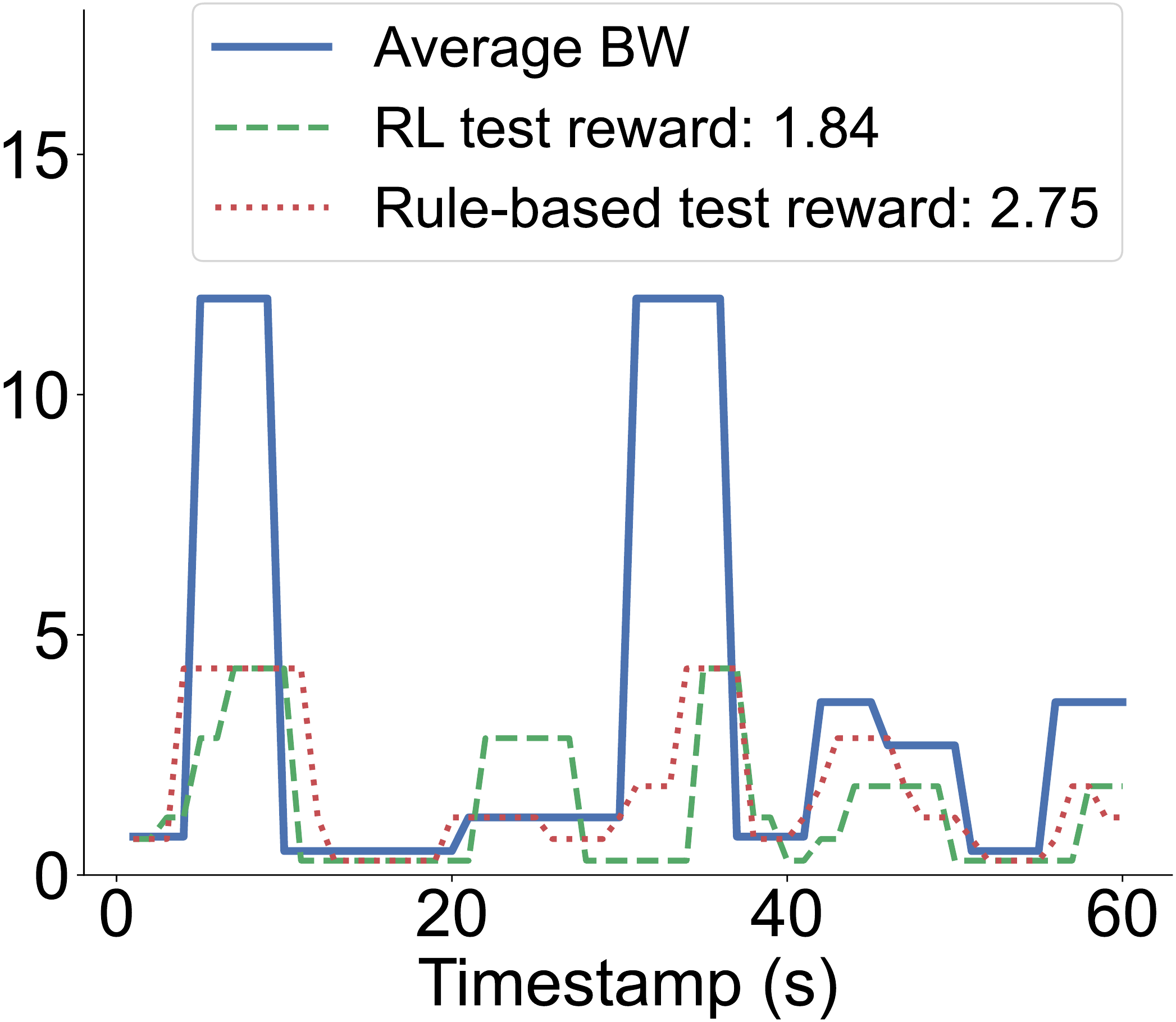}
        \caption{Trace in Y (improvable)}
        \label{fig:trace-example-z}
    }
    \end{subfigure}
    \vspace{1pt}
    \tightcaption{Contrasting (a) an inherently hard (possibly unsolvable) environment with (b) an improvable environment. 
    The difference is that the rule-based policy's reward is higher than the RL policy in (b), whereas their rewards are similar in (a).}
    \label{fig:difficulty-example}
\end{figure}

\mypara{Example}
\zx{Figure~\ref{fig:motivation-optimal} shows a concrete example in ABR, where ``Strawman 3'' leads to a suboptimal outcome. 
(\S\ref{subsec:eval:micro} will empirically test these three strawman approaches.)
We first pretrain an RL-based ABR policy which performs poorly on $X$ and $Y$ (two sets of bandwidth traces from two different environment configurations, details in \S\ref{app:trace-adding}).
\zxedit{Since the performance gap between the current RL model and the optimum is larger on $X$ than on $Y$, Strawman 3 opts for adding $X$  to the training in the next step.}
However, Figure~\ref{fig:motivation-optimal} shows that training further on $X$ yields only a marginal reward improvement on $X$ (and also hurts the performance on $Y$).

Instead, adding $Y$ to training is a better choice at this point---the performance on $Y$ is significantly improved (and it also benefits the performance on $X$ though to a less extent). 

To take a closer look, we plot two example traces from $X$ and $Y$ in Figure~\ref{fig:difficulty-example}:
The trace from $X$ fluctuates with a smaller magnitude but more frequently, whereas the trace from $Y$ fluctuates with a greater magnitude but much less frequently. 
However, such observations cannot generalize to an arbitrary pair of environments or a different application.
}

\section{Design and implementation of \name}
\subsection{Curriculum generation}
\label{subsec:design:curricula}

To identify rewarding environments, the idea of \name is to find environments with a large {\em gap-to-baseline}, \ie the RL policy is worse than a given rule-based baseline by a large margin.
At a high level, adding such environments to training has three practical benefits. 

\begin{figure}[t!]
    \centering
    \begin{subfigure}[t]{0.99\linewidth}
    {
        \includegraphics[width=\linewidth]{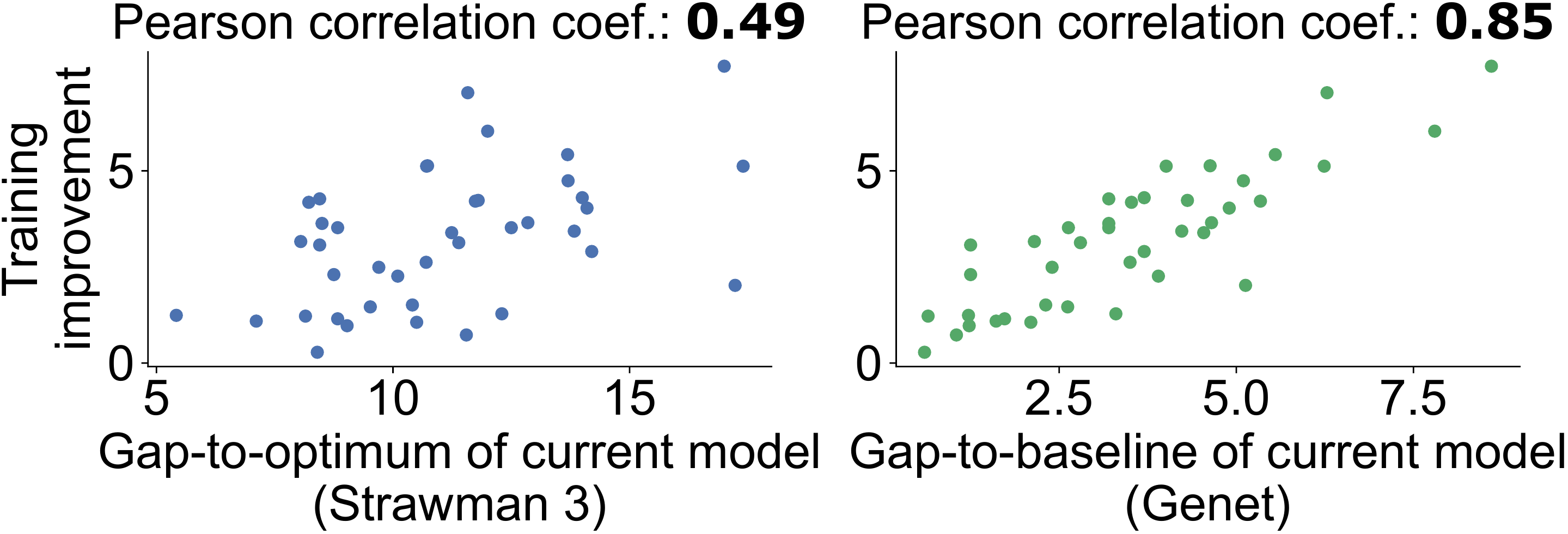}
        \vspace{-15pt}
        \caption{ABR}
        \label{fig:}
        \vspace{5pt}
    }
    \end{subfigure}
    \begin{subfigure}[t]{0.99\linewidth}
    {
        \includegraphics[width=\linewidth]{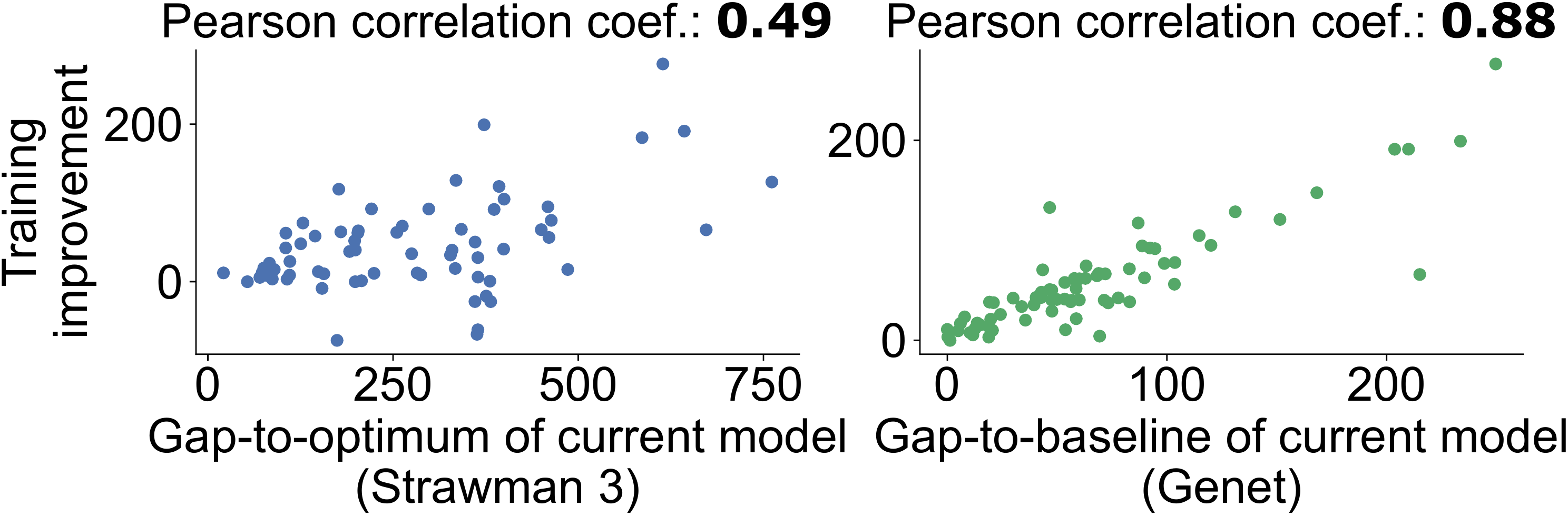}
        \vspace{-15pt}
        \caption{CC}
        \label{fig:}
    }
    \end{subfigure}
    \tightcaption{Compared with the \zxedit{gap-to-optimum} (left), the current model's gap-to-baseline (right) in an environment is more indicative of its potential training improvement in the environment.}
    \vspace{-3pt}
    \label{fig:scatter}
\end{figure}

First, when a rule-based baseline performs much better than the RL policy in an environment, it means that
the RL model may learn to ``imitate'' the baseline's known rules while training in the environment, bringing it on par with---if not better than---the baseline. \footnote{This may not be true when the behavior of the rule-based algorithm cannot be approximated by RL's policy DNN, and we will discuss this issue in \S\ref{sec:discuss}.}
Therefore, a large gap-to-baseline indicates plausible room for the current RL model to improve. 
Figure~\ref{fig:scatter} empirically confirms this with one example ABR policy and CC policy (both are intermediate models during \name-based training).
For example, among 73 randomly chosen synthetic environment configurations in CC, a configuration with a larger gap-to-baseline is likely to yield more improvement when adding its environments to the RL training.
Moreover, this correlation is stronger than using the \zxedit{performance gap between the current model and the optimum} (``Strawman 3'' in \S\ref{sec:curricula}) to decide which environments are rewarding. 
\zx{Nonetheless, the model's training improvement does not only depend on the gap-to-baseline. Other factors such as training hyperparameters can affect the reward improvement of an RL model. For example, too large a learning rate causes the RL model to jump over the optima while too small a learning rate slows down the convergence. In this work, we only focus on the gap-to-baseline and keep the training hyperparameters (\eg learning rate, batch size of each iteration) unchanged in all the experiments.}

Second, although not all rule-based algorithms are easily interpretable or completely fail-proof, many of them have traditionally been used in networked systems long before the RL-based approaches and are considered more trustworthy than black-box RL algorithms. 
Therefore, operators tend to scrutinize any performance disadvantages of the RL policy compared with the rule-based baselines currently deployed in the system.
By promoting environments with large gap-to-baselines, \name directly reduces the possibility that the RL policy causes performance regressions.

In short, the gap-to-baseline builds on the insight that rule-based baselines are {\em complementary} to RL policies---they are less susceptible to any discrepancies between training and test environments, whereas the performance of an RL policy is potentially sensitive to the environments seen during training.
In \S\ref{subsec:eval:micro}, we will discuss the impact of different choices of rule-based baselines and why gap-to-baseline is a better way of using the rule-based baseline than alternatives.
It is worth noting that the rewarding environments (those with large gap-to-baselines) 
do {\em not} have particular meanings outside the context of a given pair of RL model and baseline. 
For instance, when an RL-based CC model has a greater gap-to-baseline in some network environments, it only means that it is easier to improve the RL model by training it in these environments; it does not indicate if these environments are easy or challenging to any traditional CC algorithm.

\subsection{Training framework}
\label{subsec:design:framework}

\begin{figure}[t]
    \centering
    \includegraphics[width=0.99\columnwidth]{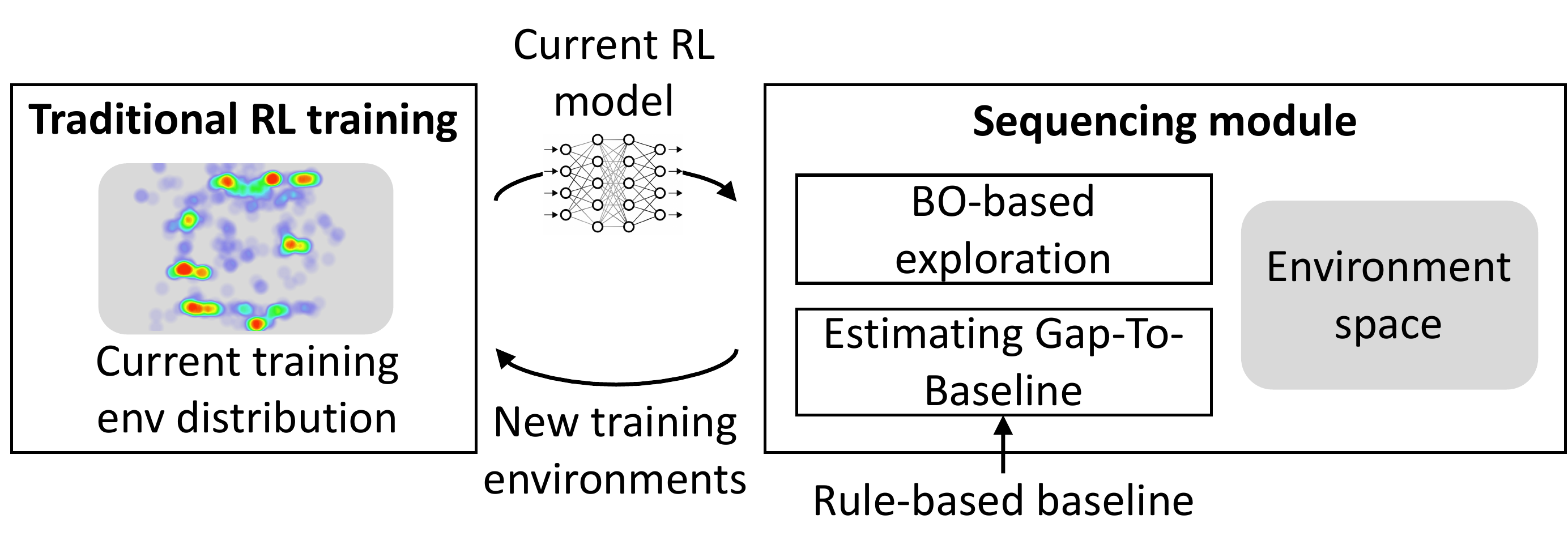}
    \vspace{-4pt}
    \tightcaption{Overview of \name's training process.}
    \label{fig:diagram}
    \vspace{-2pt}
\end{figure}

Figure~\ref{fig:diagram} depicts \name's high-level iterative workflow to realize curriculum learning. 
Each iteration consists of three steps (which will be detailed shortly):
\begin{packedenumerate}
\item First, we update the current RL model for a fixed number of iterations over the current training environment distribution; 
\item Second, we select the environments where the current RL model has a large gap-to-baseline; and
\item Third, we promote these selected environments in the training environments distribution used by the RL training process in the next iteration. 
\end{packedenumerate}

\mypara{Training environment distribution}
We define a distribution of training environments as a probability distribution over the space of {\em configurations}, each being a vector of 5--6 parameters (summarized in Table~\ref{tab:ABR-params}, \ref{tab:CC-params}, \ref{tab:LB-params}) used to generate network environments. An example configuration is: [BW: 2--3Mbps, BW changing frequency: 0--20s, Buffer length: 5--10s].
\name sets the initial training environment distribution to be a uniform or exponential distribution along with each parameter, and automatically updates the distribution used in each iteration, effectively generating a training curriculum.

When recorded traces are available, \name can augment the training with
trace-driven environments as follows.
Here we use bandwidth traces as an example. 
The first step is to categorize each bandwidth trace along with the bandwidth-related parameters (\ie bandwidth range and variance in our case). 
Each time a configuration is selected by RL training to create new environments,
with a probability of $w$ (30\% by default),
\name samples a bandwidth trace whose bandwidth-related parameters fall into the range
of the selected configuration.

In \S\ref{subsec:eval:perf}, we will show that adding trace-driven environments to training improves the performance of RL policies, especially when tested in unseen real traces from the same distribution. 
That said, even if we do not use trace-driven environments in RL training, our trained RL policies still outperform the traditional method of training RL over real traces or synthetic traces.

\mypara{Key components}
Each round of \name starts with training the current model for a fixed number of iterations (defaults to 10).
Here, \name reuses the traditional training method in prior work (\ie uniform sampling of training environments per iteration), which makes it possible to incrementally apply \name to existing codebases (see our implementation in \S\ref{sec:impl}).
Recent work on domain randomization~\cite{sadeghi16, tobin17, peng18} also shows that a similar training process can benefit the generalization of RL policies~\cite{sadeghi16, tobin17, peng18}. 
The details of the training process are described in Algorithm~\ref{alg:udr}. 

After a certain number of iterations, the current RL model and a pre-determined rule-based baseline are given to a {\em sequencing module} to search for the environments where the current RL model has a large gap-to-baseline.
Ideally, we want to test the current RL model on all possible environments and identify the ones with the largest gap-to-baseline, 
but this is prohibitively expensive. 
Instead, we use {\em Bayesian Optimization}~\cite{frazier2018tutorial} (BO) as follows.
We view the expected gap-to-baseline over the environments created by configuration $\Config$ as a function of $\Config$: $Gap(\Config)=\RewardFunc(\Policy^{rule},\Config)-\RewardFunc(\Policy^{rl}_\theta,\Config)$, where $\RewardFunc(\Policy,\Config)$ is the average reward of a policy $\Policy$ (either the rule-based baseline $\Policy^{rule}$ or the RL model $\Policy^{rl}_\theta$) over $k$ (10 by default) environments randomly generated by configuration $\Config$.
BO then searches in the environment space for the configuration that maximizes $Gap(\Config)$. 

Once a new configuration is selected, the environments generated by this configuration are then added to the training distribution as follows.
When the RL training process samples a new training environment, it will choose the new configuration with $w$ probability (30\% by default) or uniformly sample a configuration from the old distribution with $1-w$ probability (70\% by default), and then create an environment based on the selected configuration.
Next, training is resumed over the new environment distribution. 

It is important to notice that the BO-based search does {\em not} carry its states when searching rewarding environments for a new RL model. 
Instead, \name restarts the BO search every time the RL model is updated.
The reason is that the rewarding environments can change once the RL model changes. 

\mypara{Design rationale}
The process described above embeds several design decisions that make it efficient. 

\vspace{0.1cm}
{\em How to choose rule-based baselines?}
For \name to be effective, the baselines should not fail in simple environments; otherwise, \name would ignore them given that the RL policy could easily beat the baselines.
For instance, when using Cubic as the baseline in training RL-based CC policies, we observe that the RL policy is rarely worse than Cubic along the dimension of random loss rate,
because Cubic's performance is susceptible to random packet losses. 
That said, we find that the choice of baselines does not significantly impact the effectiveness of \name, although a better choice tends to yield more improvement (as shown in \S\ref{subsec:eval:micro}).\footnote{One possible refinement in this regard is to use an ``ensemble'' of rule-based heuristics, and let the training scheduler focus on environments where the RL policy falls short of any one of a set of rule-based heuristics.}

\vspace{0.1cm}
{\em Why is BO-based exploration effective?}
\zx{\name models the selection of network environments that maximize gap-to-baseline as a parameter search procedure in a high-dimensional space---each dimension of the space is a configuration of the network environment (\eg link
latency), each point in the space is a set of network environments with the same configurations, and the desired points are those whose environments have large gap-to-baselines.
This problem has two features: 
(1) the environment search space is high-dimensional, and
(2) evaluating the gap-to-baseline of a point in the space is computationally expensive (partly due to the variance among the environments with the same configurations).
In this context, BO is merely one of the candidate solutions among several others to perform the parameter search. 
In \S\ref{subsec:eval:micro}, we will compare BO's efficiency with other candidate solutions and show that BO is efficient at identifying rewarding environments.} 

\zx{
\vspace{0.1cm}
{\em Why not set a threshold for the gap-to-baseline of the selected environments?}
While \name uses BO to search rewarding environments with a fixed number of steps (default is 15), an alternative is to run BO until it finds an environment configuration whose gap-to-baseline is above a threshold.
However, the latter strategy may not end (or take a long time to finish) if the RL model is already better than the baseline in most environments, which is possible during training. 
Moreover, the threshold introduces another hyperparameter to be tuned with domain knowledge.
}

\vspace{0.1cm}
{\em Impact of forgetting?}
It is important that we train models over the {\em full} range of environments.
\name does begin the training over the whole space of environment in the first iteration, but each subsequent iteration introduces a new configuration, thus diluting the percentage of random environments in training. 
This might lead to the classic problem of forgetting---the trained model may forget how to handle environments seen before. 
While we do not address this problem directly, we have found that \name is affected by this issue only mildly.
The reason is that \name stops the training after changing the training distribution for 9 times, and by then, the original environment distribution still accounts for about 10\%.\footnote{When we impose a minimum fraction of ``exploration'' (\ie uniformly randomly picking an environment from the original training distribution) in the training (which is a typical strategy to prevent forgetting~\cite{zaremba2014learning}), \name's performance becomes worse.}

\vspace{5pt}
\subsection{Implementation}
\label{sec:impl}

\name is fully implemented in Python and Bash, and has been integrated with three existing RL training codebases. 
Next, we describe the interface and implementation of \name, as well as optimizations for eliminating \name's performance bottlenecks.

\begin{figure}[t]
    \centering
    \includegraphics[width=0.99\columnwidth]{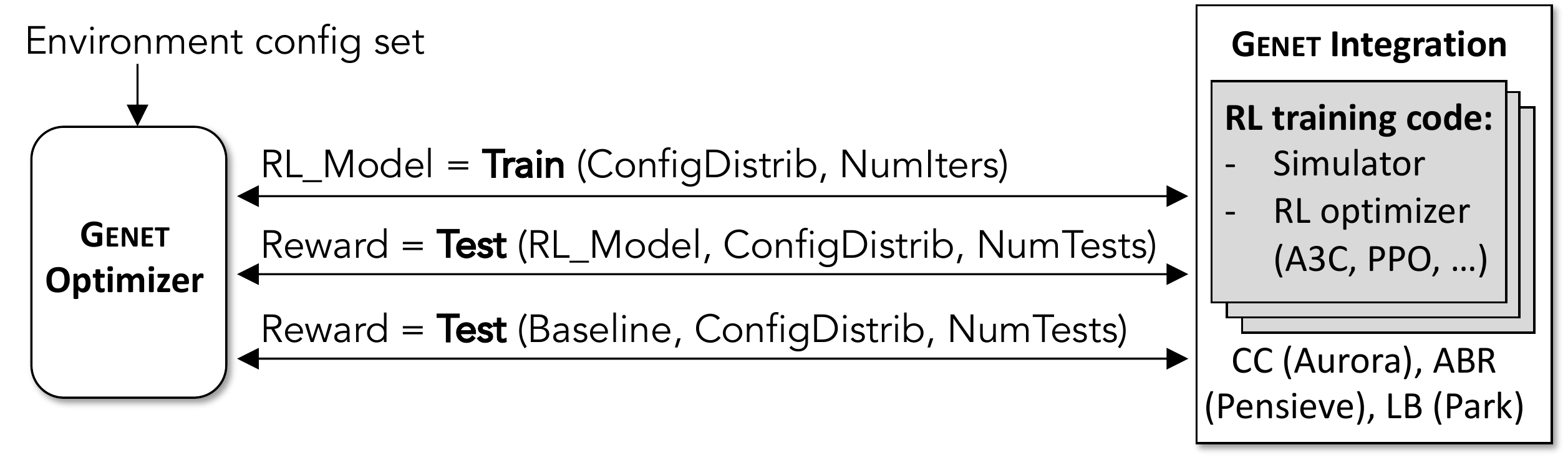}
    \tightcaption{Components and interfaces needed to integrate \name with an existing RL training codebase.}
    \label{fig:integration}
\end{figure}

\mypara{API}
\name interacts with an existing RL training codebase with two APIs (Figure~\ref{fig:integration}):
\texttt{Train} signals the RL to continue the training using the given distribution of environment configurations and returns a snapshot of the model after a specified number of training iterations;
\texttt{Test} calculates the average reward of a given algorithm (RL model or a baseline) over a specified number of environments drawn from the given distribution of configurations.

\mypara{Integration with RL training}
We have integrated \name with Pensieve ABR~\cite{pensieve-code}, Aurora CC~\cite{aurora-code}, and Park LB~\cite{park-code}, which use different RL algorithms (\eg A3C, PPO) and network simulators (\eg packet level, chunk level).
We implement the two APIs above using functionalities provided in the existing codebase.

\mypara{Rule-based baselines}
\name takes advantage of the fact that many RL training codebases (including our three use cases) have already implemented at least one rule-based baseline (\eg MPC in ABR, Cubic in CC) that runs in their simulators.
In addition, we also implemented a few baselines by ourselves, including the shortest-job-first in LB, and BBR in CC.
The implementation is generally straightforward, but sometimes the simulator (though sufficient for the RL policy) lacks crucial features for a faithful implementation of the rule-based logic. 
Fortunately, \name-based RL training merely uses the baseline to select training environments, so the consequence of having a suboptimal baseline is not considerable.


\begin{table}[t]
\footnotesize
\begin{tabular}{cccc}
\toprule
{\bf Name} & \begin{tabular}[c]{@{}c@{}}{\bf Use}\\{\bf case}\end{tabular} & \begin{tabular}[c]{@{}c@{}}{\bf Training}\\ \# traces, total length (s)\end{tabular} & \begin{tabular}[c]{@{}c@{}}{\bf Testing}\\ \# traces, total length (s)\end{tabular} \\
\midrule
FCC & ABR & 85, 105.8k & 290,  89.9k \\
Norway & ABR & 115, 30.5k & 310,  96.1k\\
Ethernet & CC & 64, 1.92k & 112,  3.35k \\
Cellular & CC & 136, 4.08k & 121,  3.64k \\
\bottomrule
\end{tabular}
\vspace{10pt}
\tightcaption{Network traces used in ABR and CC tests.}
\vspace{-14pt}
\label{tab:traces}
\end{table}

\section{Evaluation}

The key takeaways of our evaluation are:
\begin{packeditemize}
\item Across three RL use cases in networking, \name improves the performance of RL algorithms when tested in new environments drawn from the training distributions that include wide ranges of environments (\S\ref{subsec:eval:perf}). 
\item \name improves the generalization of RL performance, allowing models trained over synthetic environments to perform well even in various trace-driven environments as well as on real-world network connections  (\S\ref{subsec:eval:general}). 
\item \name-trained RL policies have a much higher chance of outperforming various rule-based baselines specified during \name-based RL training (\S\ref{subsec:eval:rule}). 
\item Finally, the design choices of \name, such as its curriculum learning strategy and BO-based search, are shown to be effective compared to seemingly natural alternatives (\S\ref{subsec:eval:micro}). 
\end{packeditemize}
Given the success of curriculum learning in other RL domains, these improvements are not particularly surprising. However, by showing for the first time that curriculum learning facilitates RL training in networking, we hope to inspire more follow-up research in this direction.

\subsection{Setup}
\label{subsec:eval:setup}

We train \name for three RL use cases in networking, using their original simulators: congestion control (CC)~\cite{aurora-code}, adaptive bitrate streaming (ABR)~\cite{pensieve-code}, and load balancing (LB)~\cite{park-code}.
As discussed in \S\ref{subsec:design:curricula}, we train and test RL policies over two types of environments.

\mypara{Synthetic environments} 
We generate synthetic environments using the parameters described in detail in \S\ref{app:trace} and Table~\ref{tab:ABR-params},\ref{tab:CC-params},\ref{tab:LB-params}. 
We choose these environment parameters to cover a variety of factors that affect RL performance.
For instance, in CC tests, our environment parameters specify bandwidth (\eg the range, variance, and how often it changes), delay, queue length, etc. 

\mypara{Trace-driven environments} 
We also use real traces for CC and ABR (summarized in Table~\ref{tab:traces}) to create trace-driven environments (in both training and testing), where the bandwidth time series are set by the real traces, but the remaining environment parameters (\eg queue length or target video buffer length) are set as in the synthetic environments. 
We test ABR policies by streaming a pre-recorded video over 290 traces from FCC broadband measurements~\cite{fccdata} (labeled ``FCC'') and 310 cellular traces~\cite{norway-data} (labeled ``Norway'').
We test CC policies on 121 cellular traces (labeled ``Cellular'') and 112 Ethernet traces (labeled ``Ethernet'') collected by the Pantheon platform~\cite{pantheon}.

\mypara{Baselines}
We compare \name-trained policies with several baselines. 
First, {\em traditional RL} trains RL policies by uniformly sampling environments from the target distribution per iteration.
We train three types of RL policies (RL1, RL2, RL3) over fixed-width uniform
distribution of synthetic environments, specified in
Table~\ref{tab:ABR-params},~\ref{tab:CC-params},~\ref{tab:LB-params}. 
From RL1 to RL3, the sizes of their training environment ranges are in ascending order.

We also train RL policies over trace-driven environments, \ie randomly picking bandwidth traces from one of the recorded sets. 
This is the same as prior work, except that we also vary non-bandwidth-related parameters (\eg queue length, buffer length, video length, etc) to increase its robustness.
In addition, we test an early attempt to improve RL~\cite{robustifying} which generates new training bandwidth traces that maximize the gap between the RL policy and optimal adaptation with a non-smoothness penalty (\S\ref{subsec:eval:micro}). 

Second, {\em traditional rule-based algorithms} include BBA~\cite{bba} and RobustMPC~\cite{mpc} for ABR, PCC-Vivace~\cite{vivace}, BBR~\cite{bbr} and CUBIC for CC, and least-load-first (LLF) for LB.\footnote{By default, we use RobustMPC as MPC and PCC Vivace-latency as Vivace, since they appear to perform better than their perspective variants.}
They can be viewed as a reference point for traditional non-ML solutions.

\subsection{Asymptotic performance}
\label{subsec:eval:perf}

We first compare \name-trained policies and traditionally trained RL policies, in terms of their {\em asymptotic performance} (\ie test performance over new test environments drawn independently from the training distribution). 
In other words, we train RL policies over environments from the target distribution and test them in new environments from the same distribution.

\begin{figure}[t]
    \centering
    \includegraphics[width=0.9\linewidth]{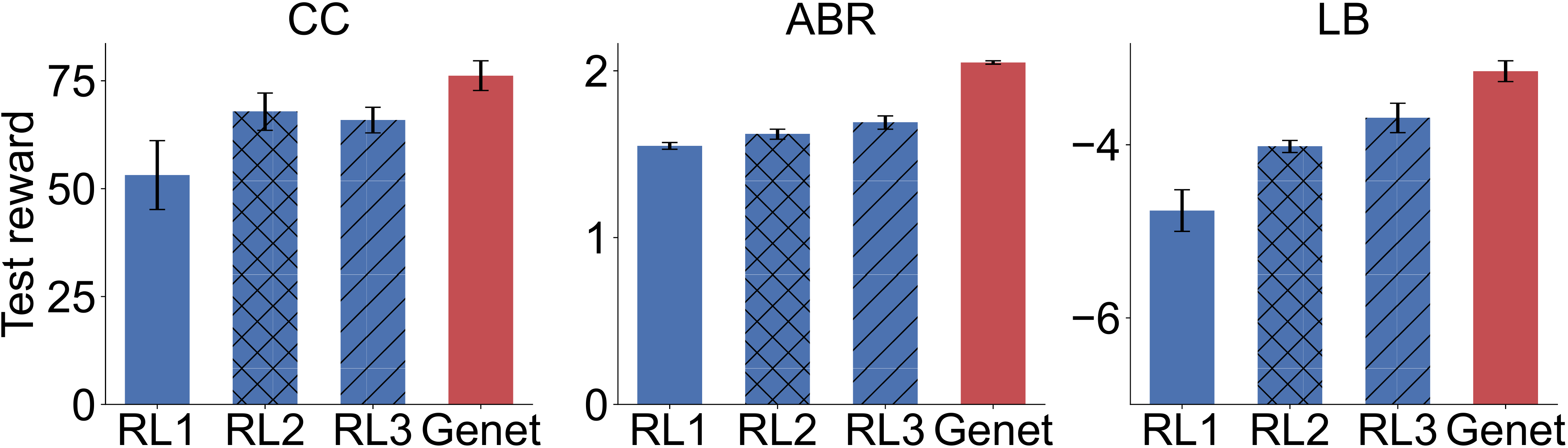}
    \vspace{2pt}
    \tightcaption{Comparing the performance of \name-trained RL policies for CC, ABR, and LB, with baselines in unseen synthetic environments drawn from the training distribution, which sets all environment parameters to their full ranges.}
    \label{fig:eval:asymptotic-synthetic}
\end{figure}

\mypara{Synthetic environments}
We first test \name-trained CC, ABR, and LB policies under their perspective RL3 synthetic ranges (where all parameters are set to their full ranges) as the target distribution. 
As shown in Figure~\ref{fig:asymptotic}, in these training ranges, traditional RL training yields little performance improvement over the rule-based baselines.
Figure~\ref{fig:eval:asymptotic-synthetic} compares \name-trained CC, ABR, and LB policies with their respective baselines over 200 new synthetic environments randomly drawn with the target distribution. 

Across three use cases, we can see that \name consistently improves over traditional RL-trained policies by
8--25\% for ABR, 14--24\% for CC, 15\% for LB, compared with traditional RL training methods.
We notice that there is no clear ranking among the three traditional RL-trained policies.
This is because RL1 helps training to converge better but only sees a small slice of the target distribution, whereas RL3 sees the whole distribution but cannot train a good model. 
In contrast, \name outperforms them, as curriculum learning allows it to learn more efficiently from the large target distribution.

\begin{figure}[t]
    \includegraphics[width=0.85\linewidth]{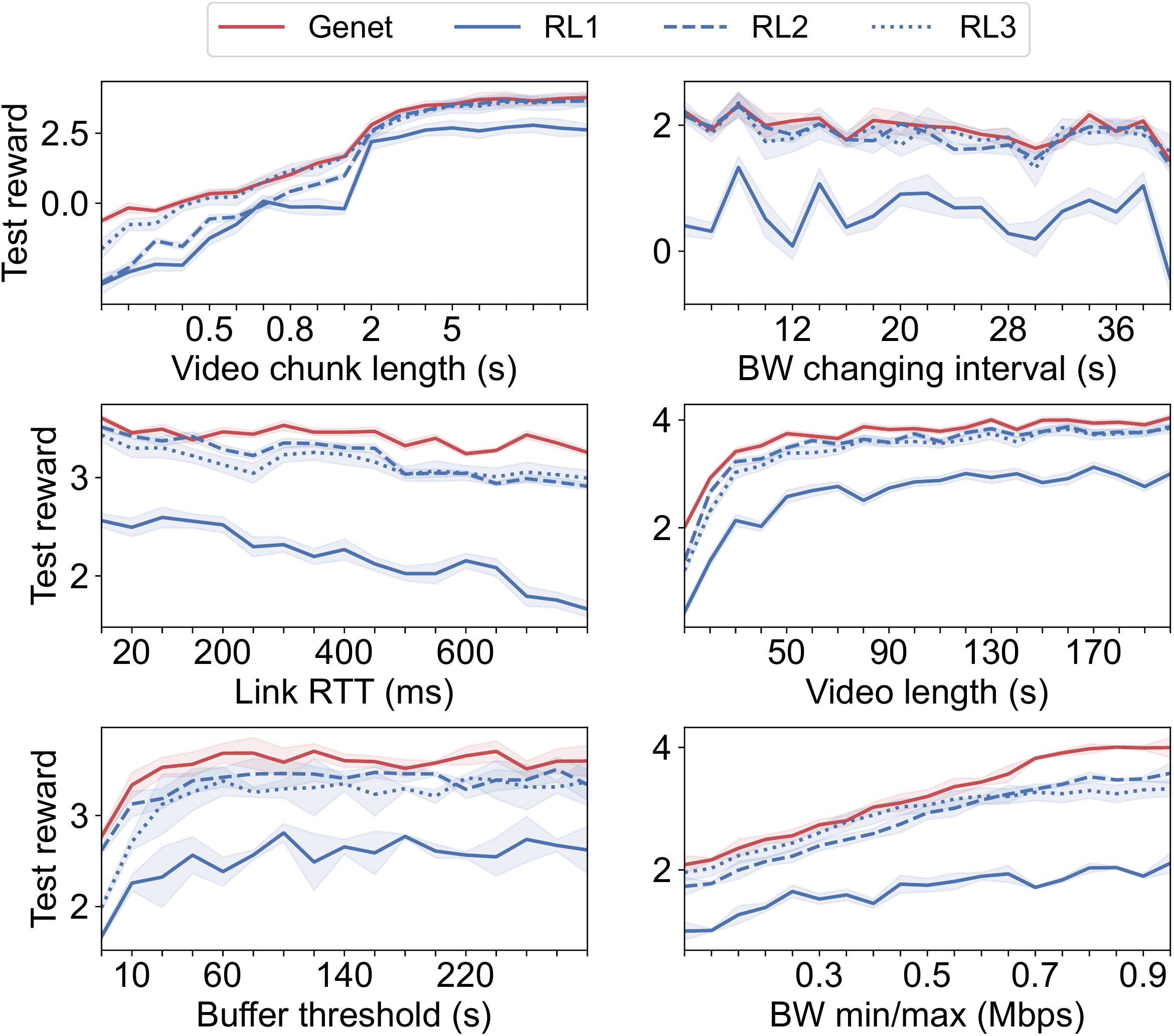}
    \tightcaption{Test of ABR policies along individual env-parameters.}
    \label{fig:eval:synthetic-setting-abr}
\end{figure}

\begin{figure}[t]
    \centering
    \includegraphics[width=0.85\linewidth]{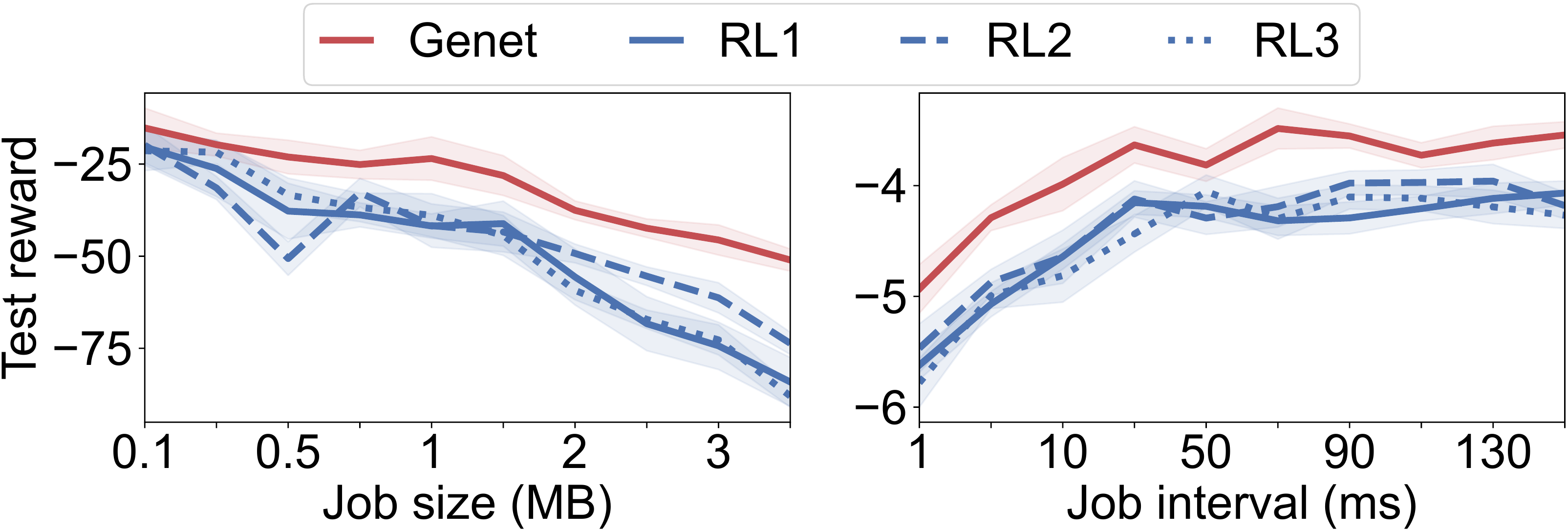}
    \tightcaption{Test of LB policies along individual env-parameters.}
    \label{fig:eval:emu-lb}
\end{figure}

To show the performance more thoroughly, Figure~\ref{fig:eval:synthetic-setting-abr} picks ABR as an example and shows the performance across different values along with six environment parameters.
We vary one parameter at a time while fixing other parameters at the same default values (see Table~\ref{tab:ABR-params}, \ref{tab:CC-params}, \ref{tab:LB-params}). 
We see that \name-trained RL policies enjoy consistent performance advantages (in reward) over the RL policies trained by traditional RL-trained models.
This suggests that the improvement of \name shown in Figure~\ref{fig:eval:asymptotic-synthetic} is not a result of improving rewards in some environments at the cost of degrading rewards in others; instead, \name improves rewards in most cases. Figure~\ref{fig:eval:emu-lb} shows that in the simulated environments \cite{park-code}, the \name-trained LB policy outperforms its baselines by 15\%.

\begin{figure}[t]
    \centering
    \begin{subfigure}[t]{0.48\linewidth}
    {
        \includegraphics[width=1\linewidth]{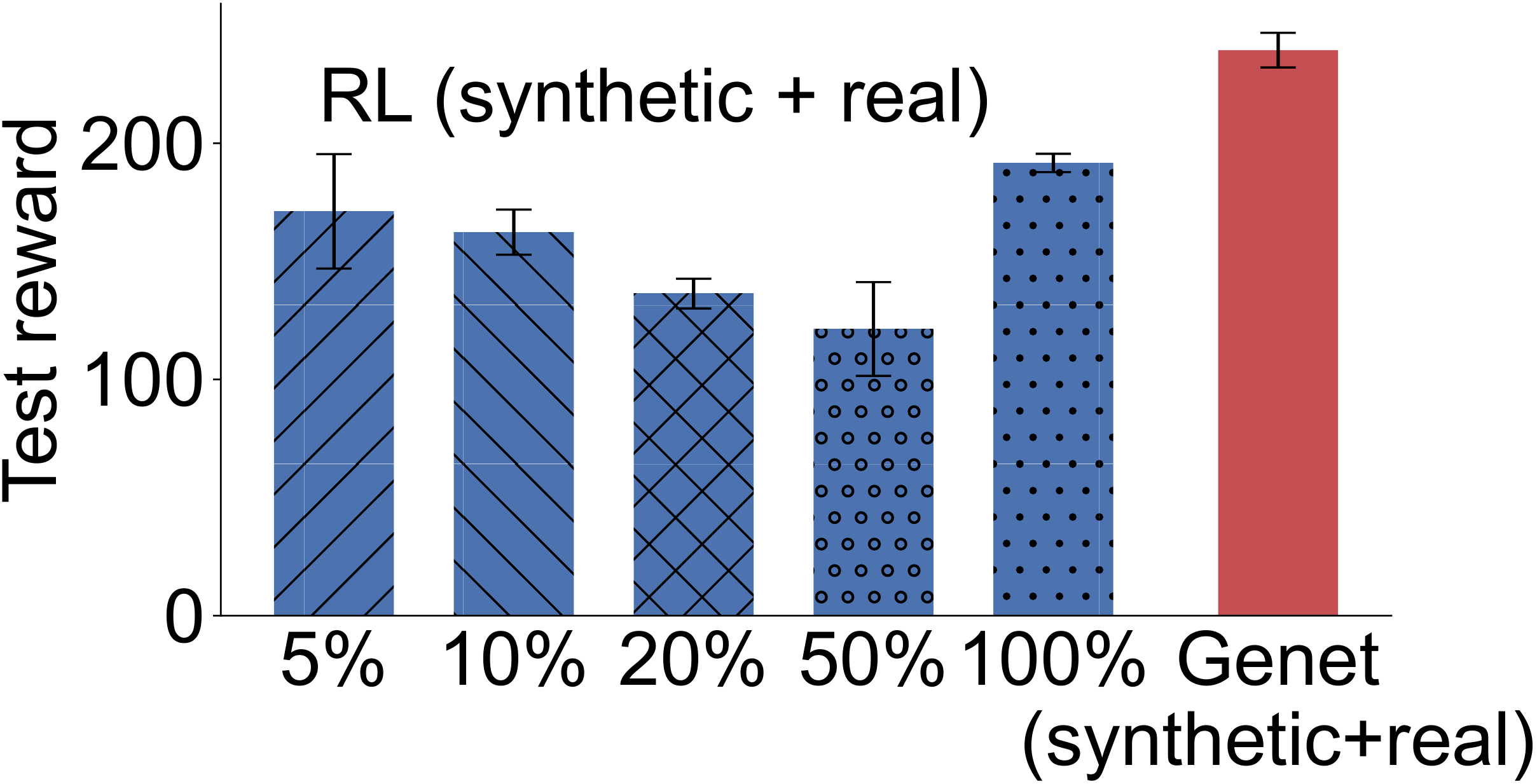}
        \caption{Congestion control (CC)}
        \label{}
    }
    \end{subfigure}
    \begin{subfigure}[t]{0.48\linewidth}
    {
        \includegraphics[width=1\linewidth]{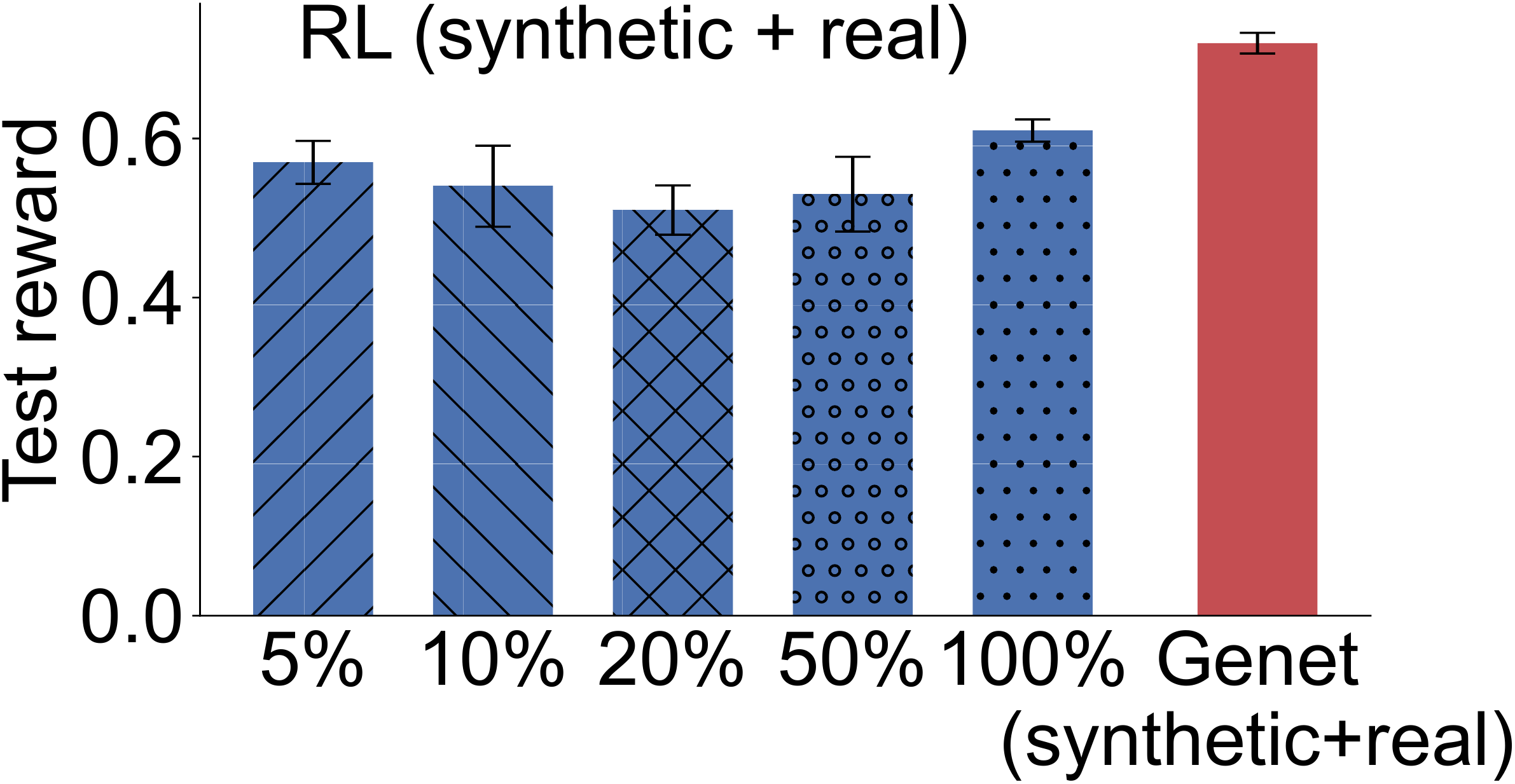}
        \caption{Adaptive bitrate (ABR)}
        \label{}
    }
    \end{subfigure}
    \tightcaption{Asymptotic performance of \name-trained CC policies (a) and ABR policies (b) and baselines, when the real network traces are randomly split into a training set and a test set.}
    \label{fig:eval:asymptotic-real}
\end{figure}

\mypara{Trace-driven environments}
Next, we set the target environment distributions of ABR and CC to be the environments generated from multiple real-world trace sets (FCC and Norway for ABR, Ethernet and Cellular for CC).
We partition each trace set as listed in Table~\ref{tab:traces}.
\name trains ABR and CC policies by combining trace-driven environments and synthetic environments (described in \S\ref{subsec:design:framework}). 
For a thorough comparison, both \name and the traditional RL training have access to the training portion of the real traces as well as the synthetic environments. 
We vary the ratio of real traces and synthetic environments and feed them to the traditional RL training method, \eg if the ratio of real traces is 20\%, then the traditional RL training randomly draws a trace-driven environment with 20\% probability and synthetic environments with 80\% probability.
That is, we test different ways for the traditional RL training to combine the training traces and synthetic environments. 
Figure~\ref{fig:eval:asymptotic-real} tests \name-trained ABR and CC policies with their respective traditional RL-trained baselines over new environments generated from the traces in the testing set. 
Figure~\ref{fig:eval:asymptotic-real} shows that \name-trained policies
outperform traditional RL training by 17--18\%, regardless of the ratio of real traces, including when training the model entirely on real traces.

\begin{figure}[t]
    \begin{subfigure}[b]{0.48\linewidth}
    {
        \includegraphics[width=0.95\linewidth]{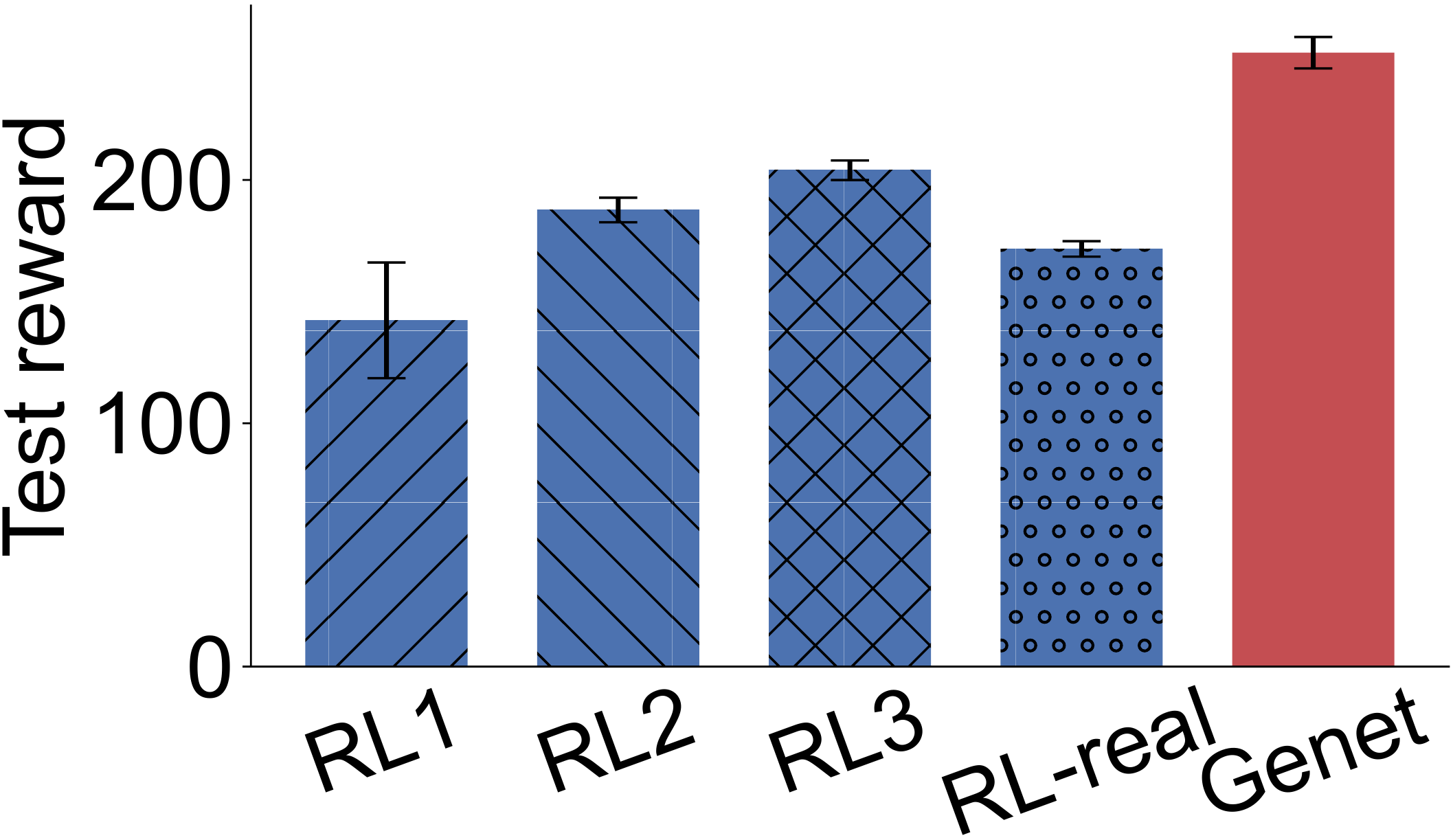}
        \caption{CC test in trace-driven environments (Cellular)}
        \label{}
    }
    \end{subfigure}
    \hfill
    \begin{subfigure}[b]{0.48\linewidth}
    {
        \includegraphics[width=0.95\linewidth]{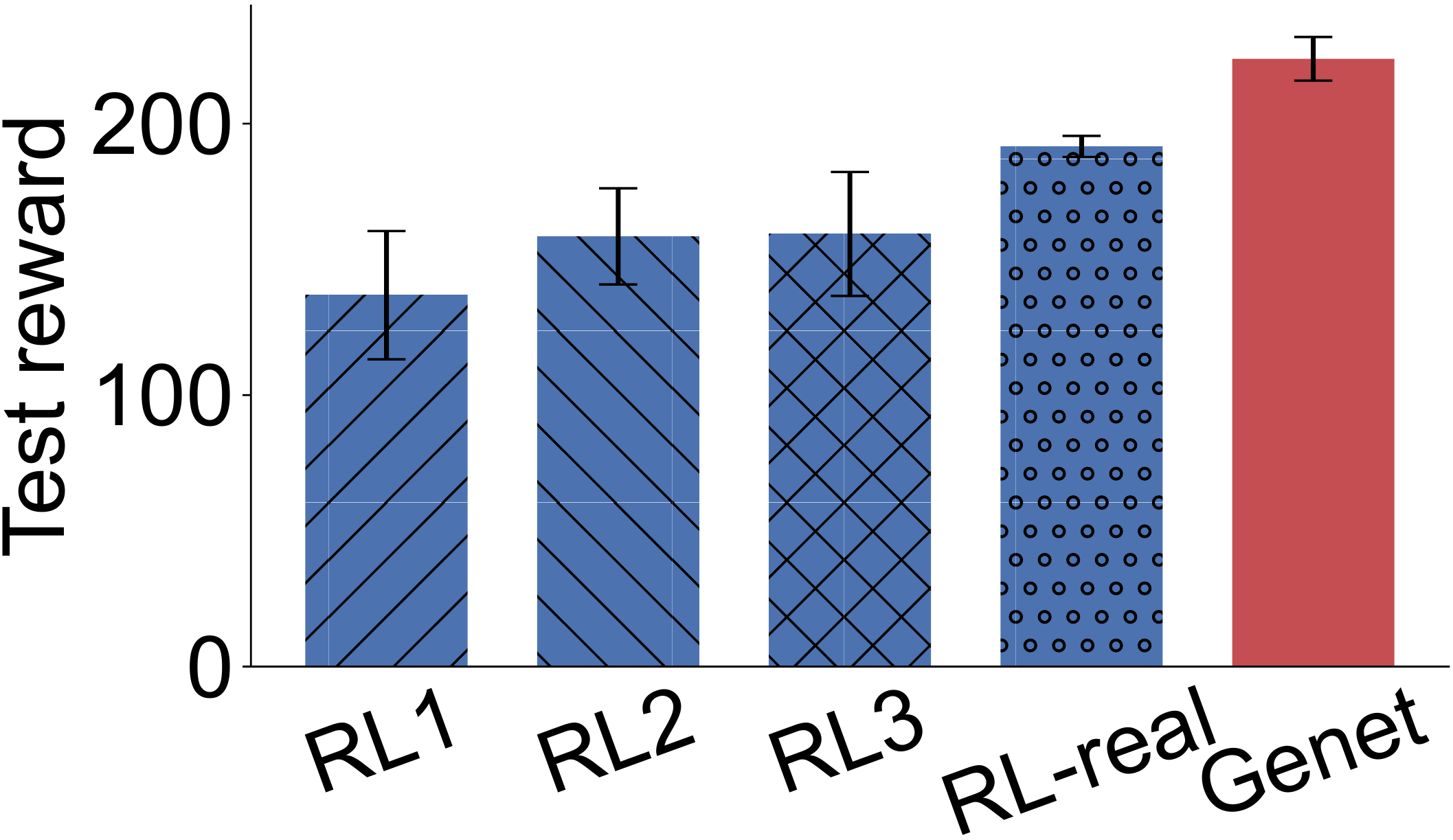}
        \caption{CC test in trace-driven environments (Ethernet)}
        \label{}
    }
    \end{subfigure}
    
    \begin{subfigure}[b]{0.48\linewidth}
    {
        \includegraphics[width=0.95\linewidth]{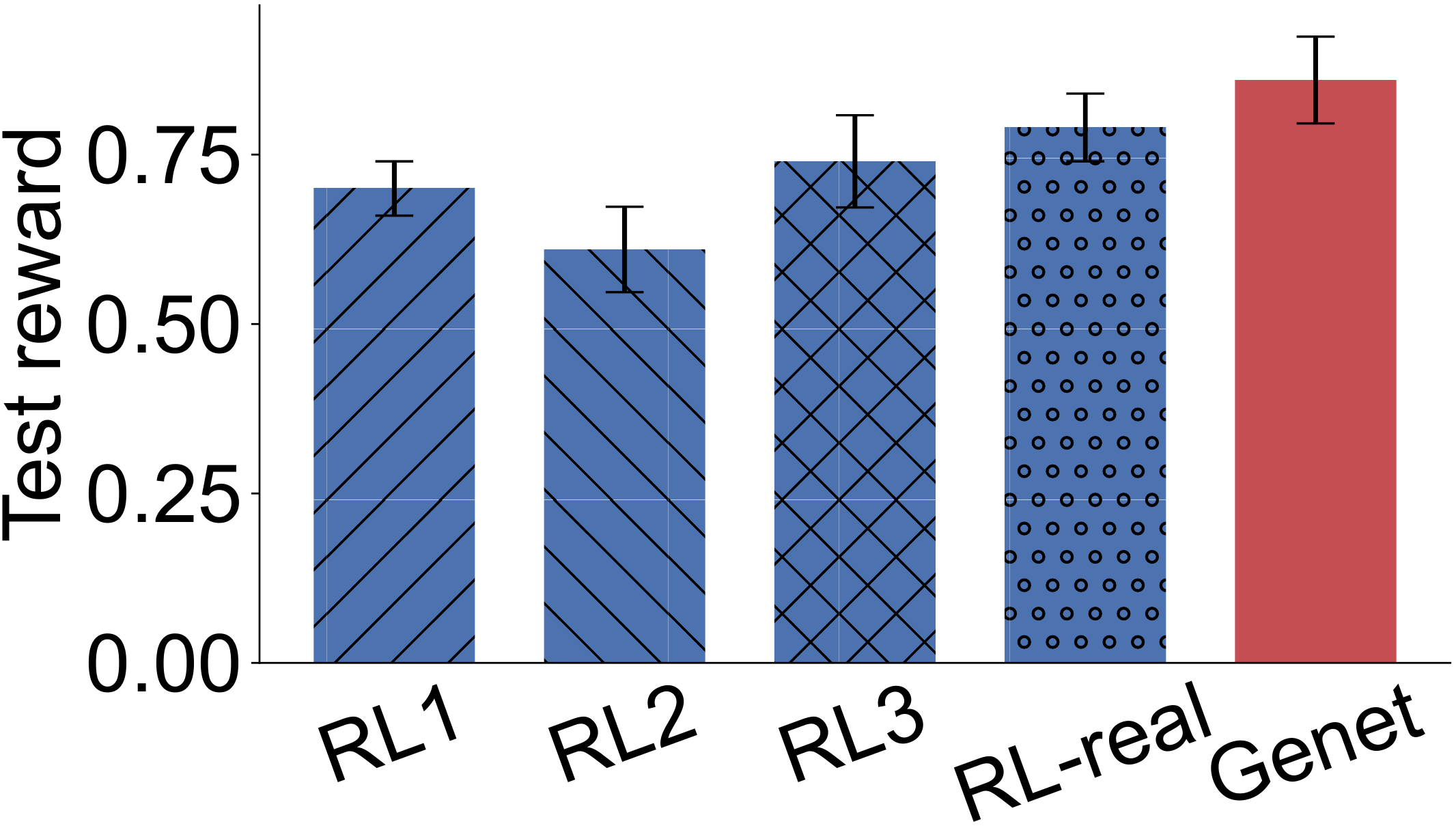}
        \caption{ABR test in trace-driven environments (FCC)}
        \label{}
    }
    \end{subfigure}
    \hfill
    \begin{subfigure}[b]{0.48\linewidth}
    {
        \includegraphics[width=0.95\linewidth]{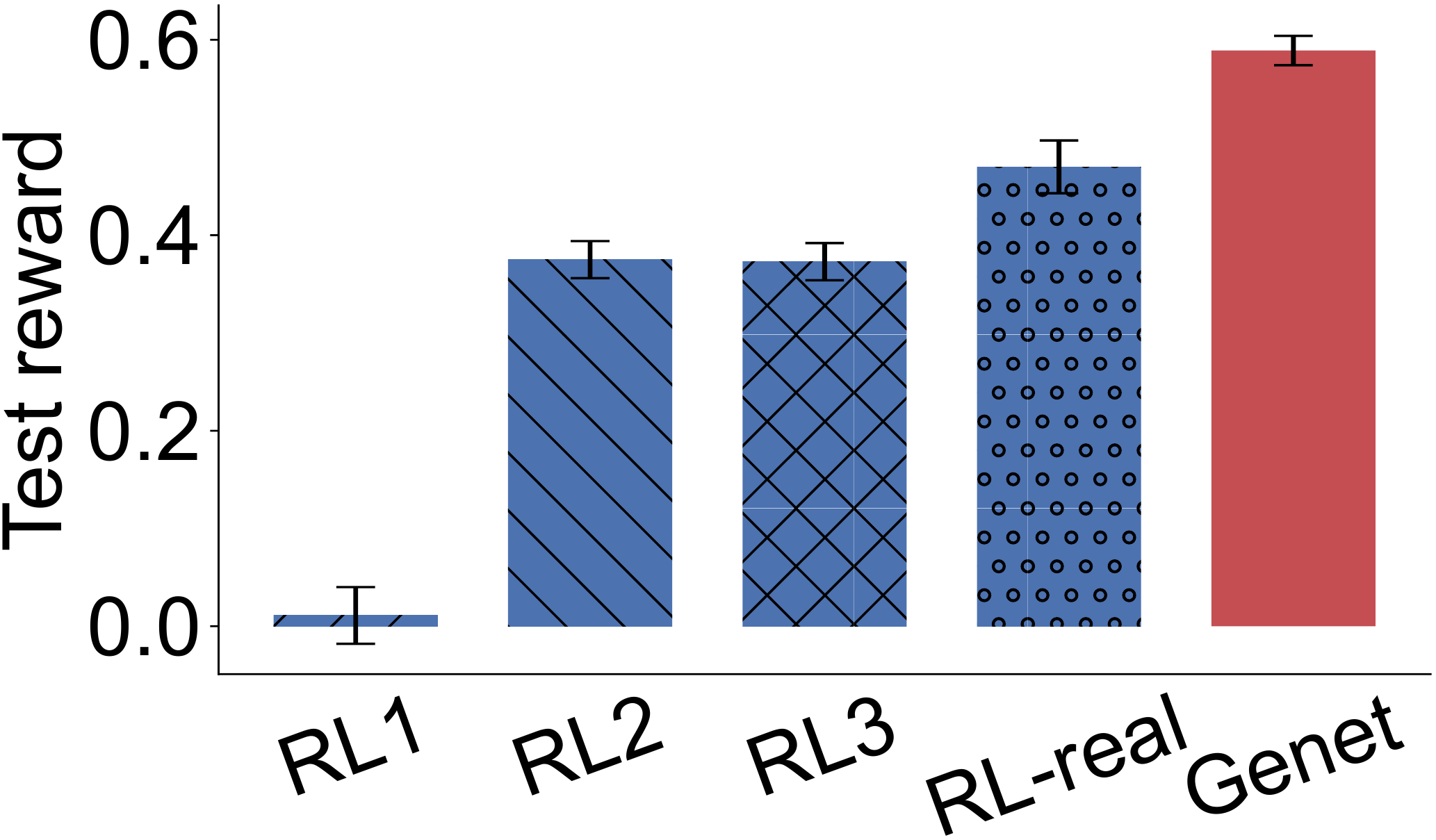}
        \caption{ABR test in trace-driven environments (Norway)}
        \label{}
    }
    \end{subfigure}
    
    \tightcaption{Generalization test: Training of various methods is done entirely in synthetic environments, but the testing is over various real network trace sets.}
    \label{fig:eval:generalization}
    \vspace{-1pt}
\end{figure}

\subsection{Generalization}
\label{subsec:eval:general}

Next, we take the RL policies of ABR and CC trained (by \name and other baselines) entirely over synthetic environments (the RL3 synthetic environment range) and test their generalization in trace-driven environments generated by the ABR (and CC) testing traces in Table~\ref{tab:traces}.

Figure~\ref{fig:eval:generalization} shows that they perform better than traditional RL baselines trained over the same synthetic environment distribution.
Though Figure~\ref{fig:eval:generalization} uses the same testing environments as Figure~\ref{fig:eval:asymptotic-real} and has a similar relative ranking between \name and traditional RL training, the implications are different: \zx{Figure~\ref{fig:eval:generalization} also shows that when the real traces are {\em not} accessible in training, \name can produce models with better generalization in real-trace-driven environments than the baselines, whereas Figure~\ref{fig:eval:asymptotic-real} shows their performance when the real traces are actively used in training of \name and the baselines.}

\subsection{Comparison with rule-based baselines}
\label{subsec:eval:rule}

\begin{figure}[t]
    \centering
    \begin{subfigure}[t]{0.45\linewidth}
    {
        \includegraphics[width=1\linewidth]{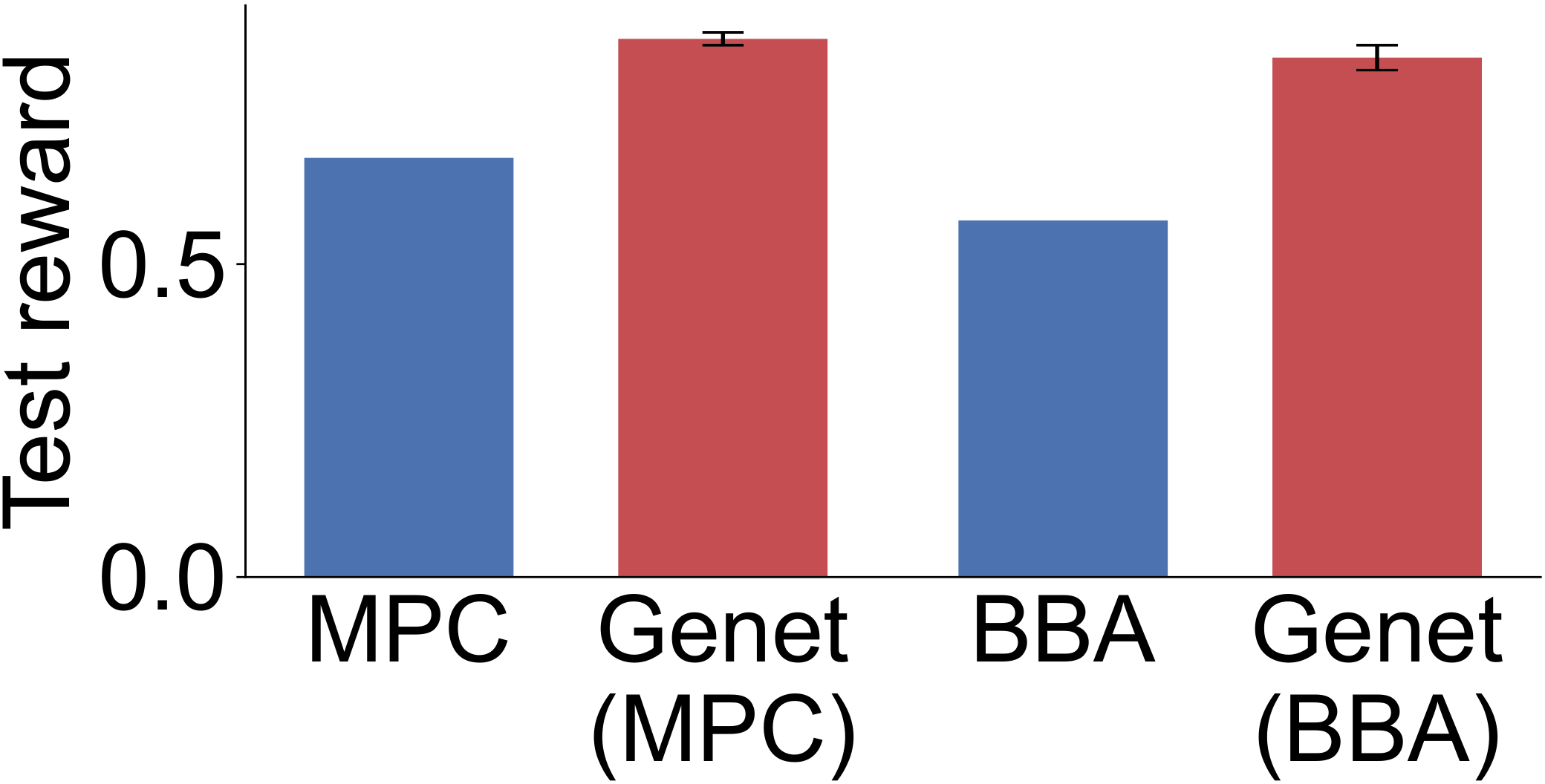}
        \caption{ABR}
        \label{}
    }
    \end{subfigure}
    \begin{subfigure}[t]{0.45\linewidth}
    {
        \includegraphics[width=1\linewidth]{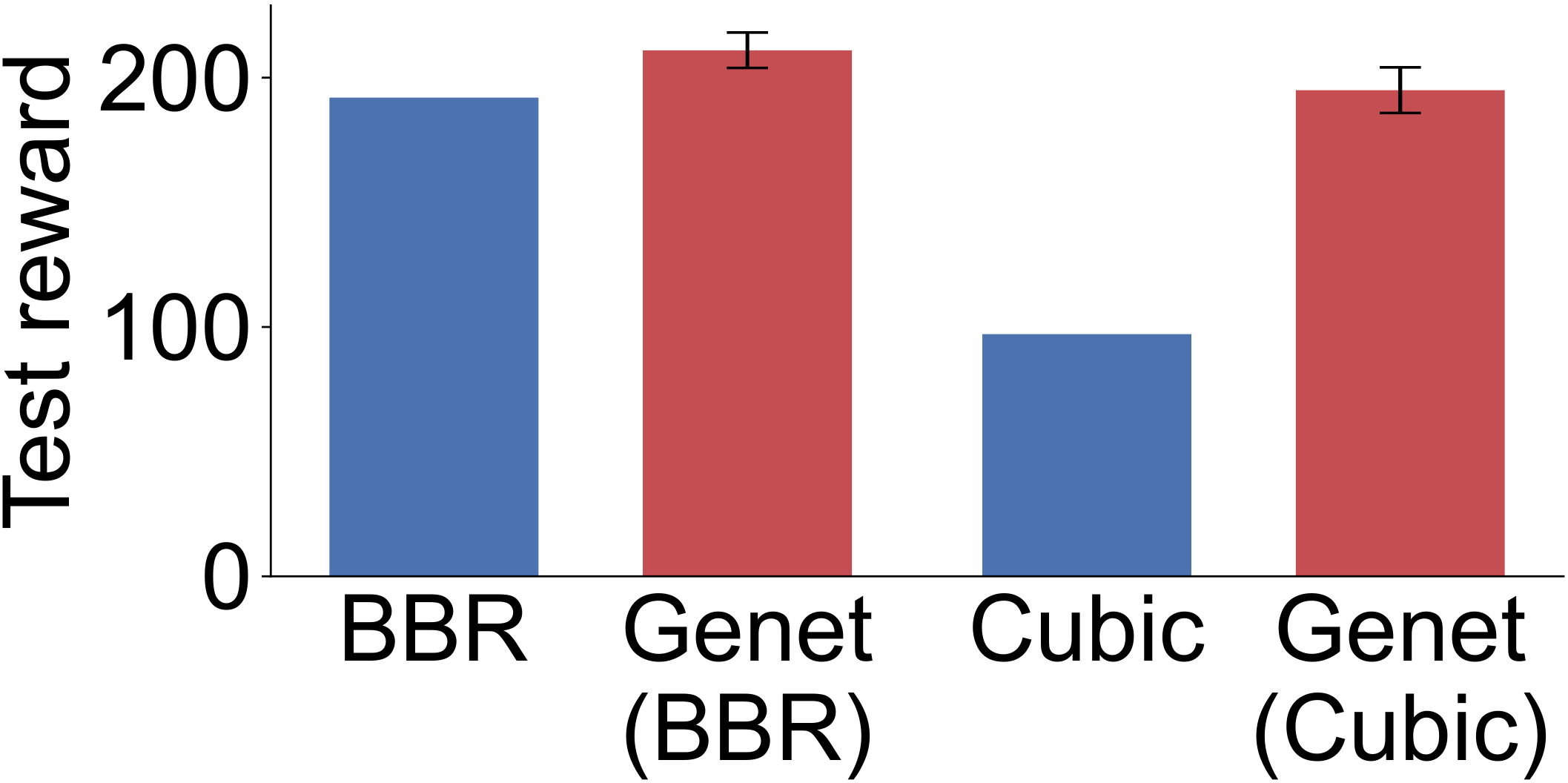}
        \caption{CC}
        \label{}
    }
    \end{subfigure}
    \tightcaption{\textit{\name outperforms the rule-based baselines used in its training.
    }}
    \vspace{-2pt}
    \label{fig:eval:baselines}
\end{figure}

\mypara{Impact of the choice of rule-based baselines}
Figure~\ref{fig:eval:baselines} shows the performance of \name-trained policies when using different rule-based baselines.
We choose MPC and BBA as baselines in the ABR experiments and BBR and Cubic as baselines in CC experiments, respectively. 
We observe that in all cases, \name-trained policies outperform their respective rule-based baselines.

\vspace{-2pt}
\begin{figure}[t]
    \centering
    \begin{subfigure}[t]{0.48\linewidth}
    {
        \includegraphics[width=0.95\linewidth]{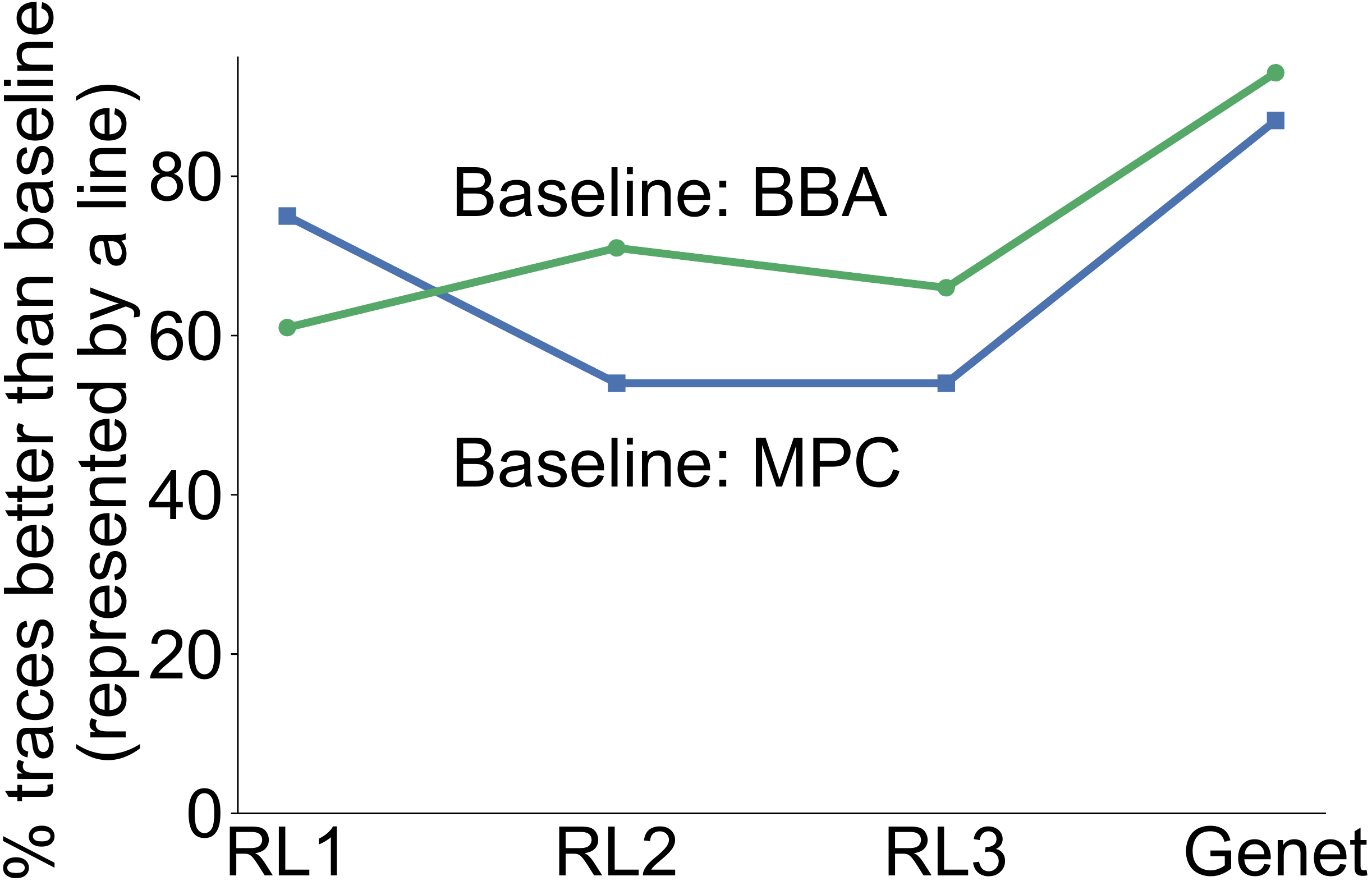}
        \caption{ABR}
        \label{}
    }
    \end{subfigure}
    \begin{subfigure}[t]{0.48\linewidth}
    {
        \includegraphics[width=0.95\linewidth]{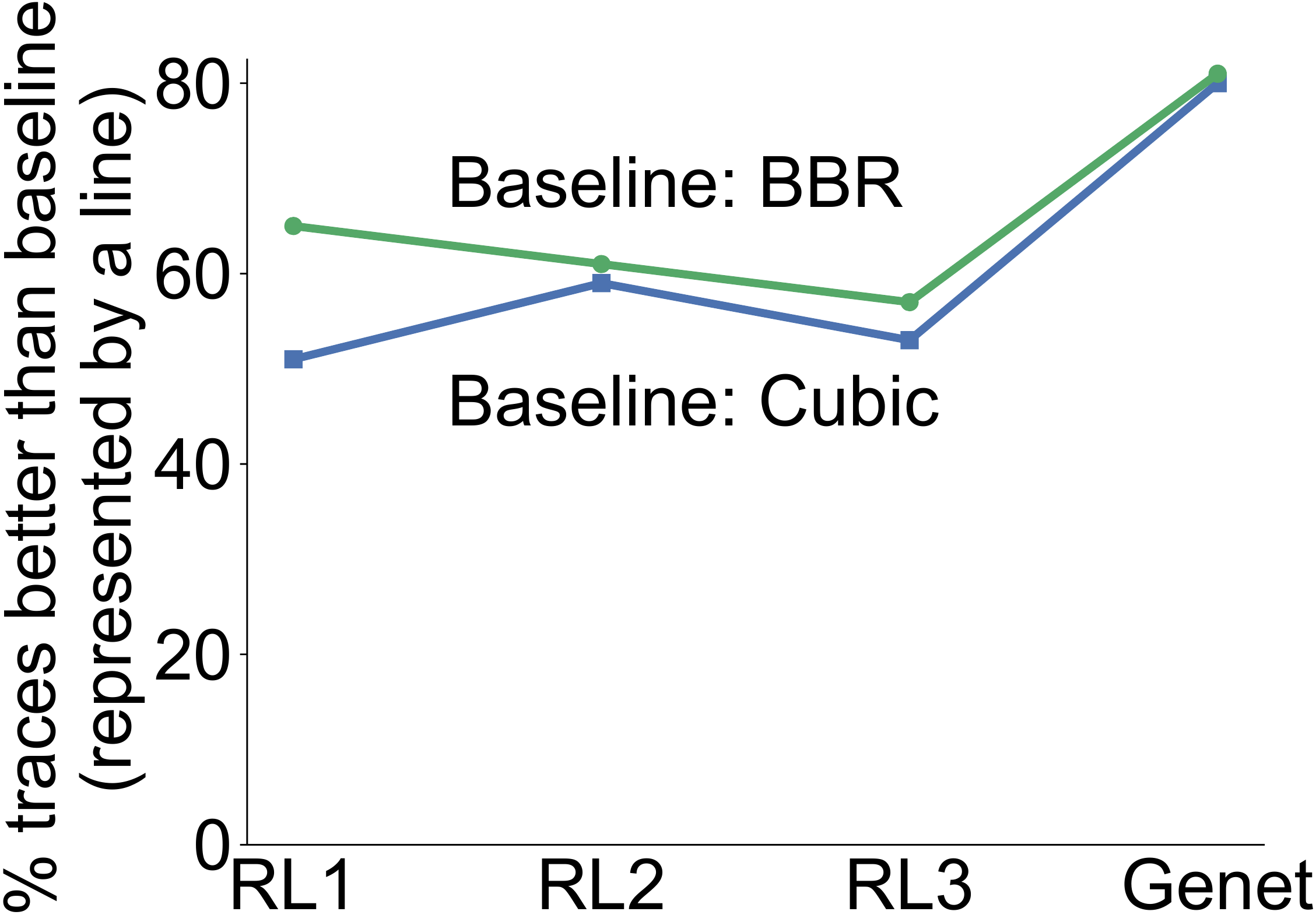}
        \caption{CC}
        \label{}
    }
    \end{subfigure}
    \tightcaption{Fraction of real traces where \name-trained policies (and traditional RL) are better than the rule-based baselines.}
    \vspace{1pt}
    \label{fig:eval:fractions}
\end{figure}

\begin{figure*}[t]
    \begin{subfigure}[b]{0.58\linewidth}
    {
        \includegraphics[height=68pt]{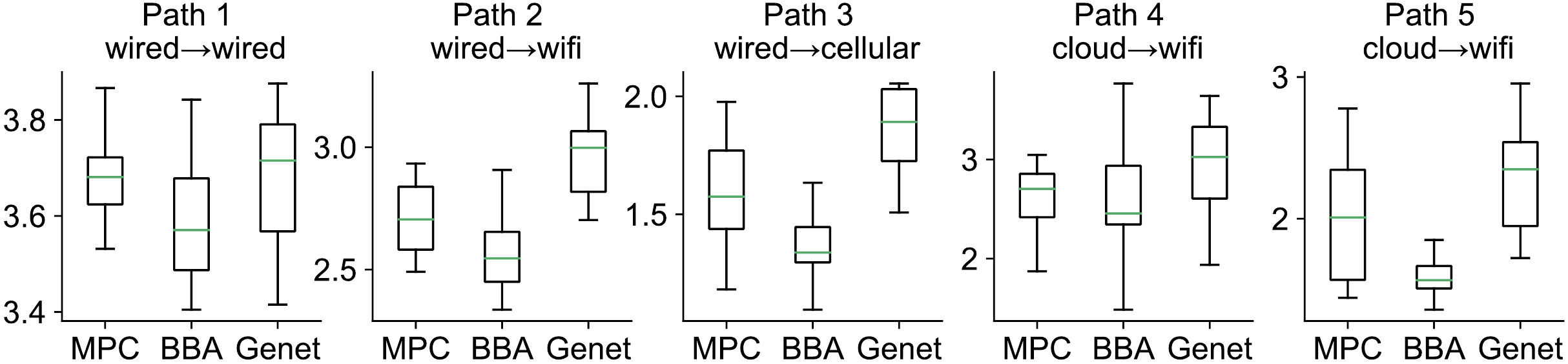}
        \caption{ABR}
        \label{fig:eval:real:abr}
    }
    \end{subfigure}
    \hfill
    \begin{subfigure}[b]{0.4\linewidth}
    {
        \includegraphics[height=68pt]{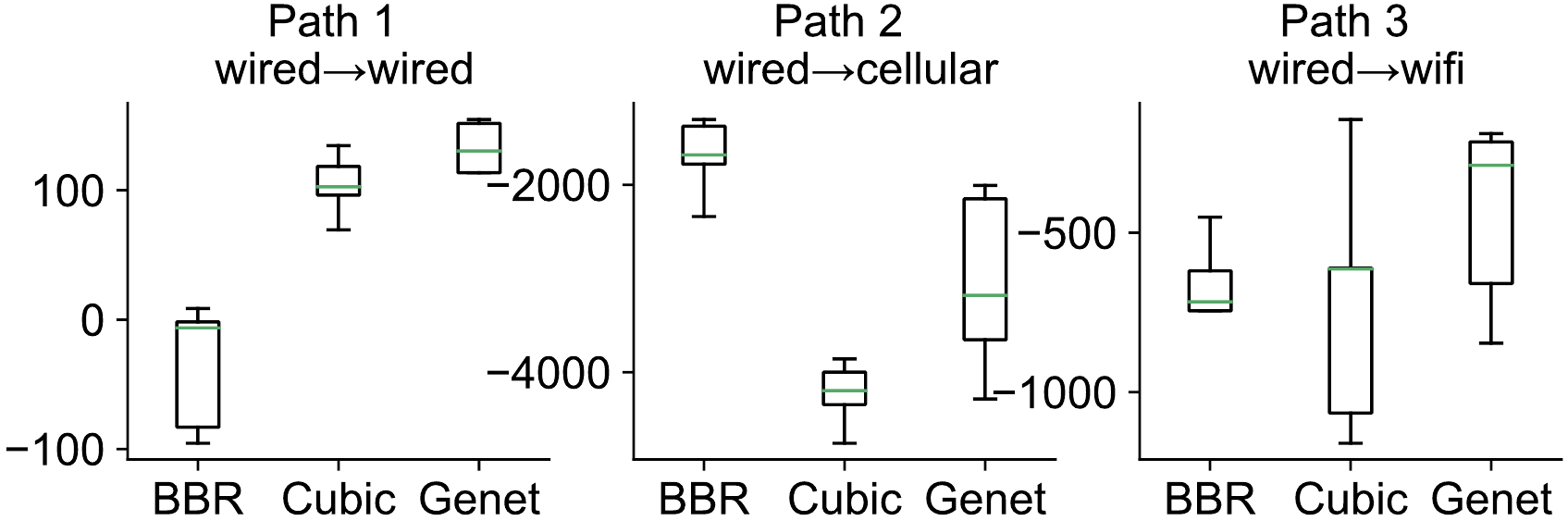}
        \caption{CC}
        \label{fig:eval:real:cc}
    }
    \end{subfigure}
    \tightcaption{Testing ABR and CC policies in real-world environments.
    }
    \label{fig:eval:real}
\end{figure*}

\myparaq{What if \name uses naive rule-based baselines}
As explained in \S\ref{subsec:design:framework},  the rule-based baseline should have a reasonable (though not necessarily optimal) performance; otherwise, it would be unable to indicate when the RL policy can be improved. 
To empirically verify it, we use two unreasonable baselines: choosing the highest bitrate when rebuffer in ABR, and choosing the highest loaded server in LB. 
In both cases, the BO-based search fails to find useful training environments, because the RL policy very quickly outperforms the naive baseline everywhere.
That said, the negative impact of using a naive baseline is restricted to the selection of training environments, rather than the RL training itself (a benefit of decoupling baseline-driven environment selection and RL training), so in the worst case, \name would be roughly as good as traditional RL training.

\myparaq{How likely is \name to outperform rule-based baselines}\\
One of \name's benefits is to increase how often the RL policy is better than the rule-based baseline used in \name.
In Figure~\ref{fig:eval:fractions}, we create various versions of \name-trained RL policies by setting the rule-based baselines to be Cubic and BBR (for CC), and MPC and BBA (for ABR).
Compared to RL1, RL2, RL3 (unaware of rule-based baselines), \name-trained policies remarkably increase the fraction of real-world traces (emulated) where the RL policy outperforms the baseline used to train them.
This suggests that operators can specify a rule-based baseline, and \name will train an RL policy that outperforms it with high probability.

\begin{figure}[t]
    \centering
    \begin{subfigure}[t]{0.48\linewidth}
    {
        \includegraphics[width=0.95\linewidth]{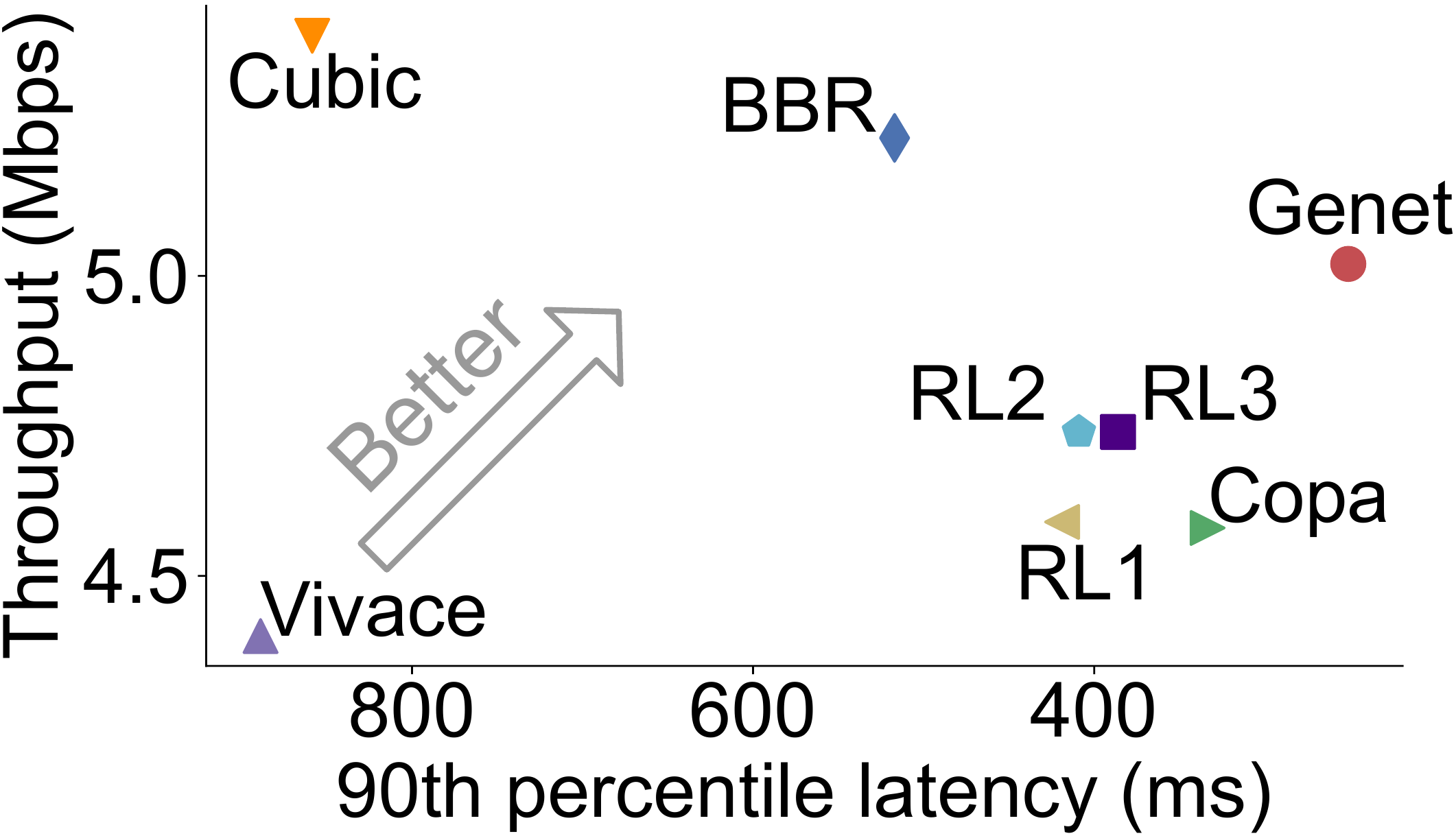}
        \caption{CC (Cellular traces)}
        \label{fig:abr:cellular}
    }
    \end{subfigure}
    \begin{subfigure}[t]{0.48\linewidth}
    {
        \includegraphics[width=0.95\linewidth]{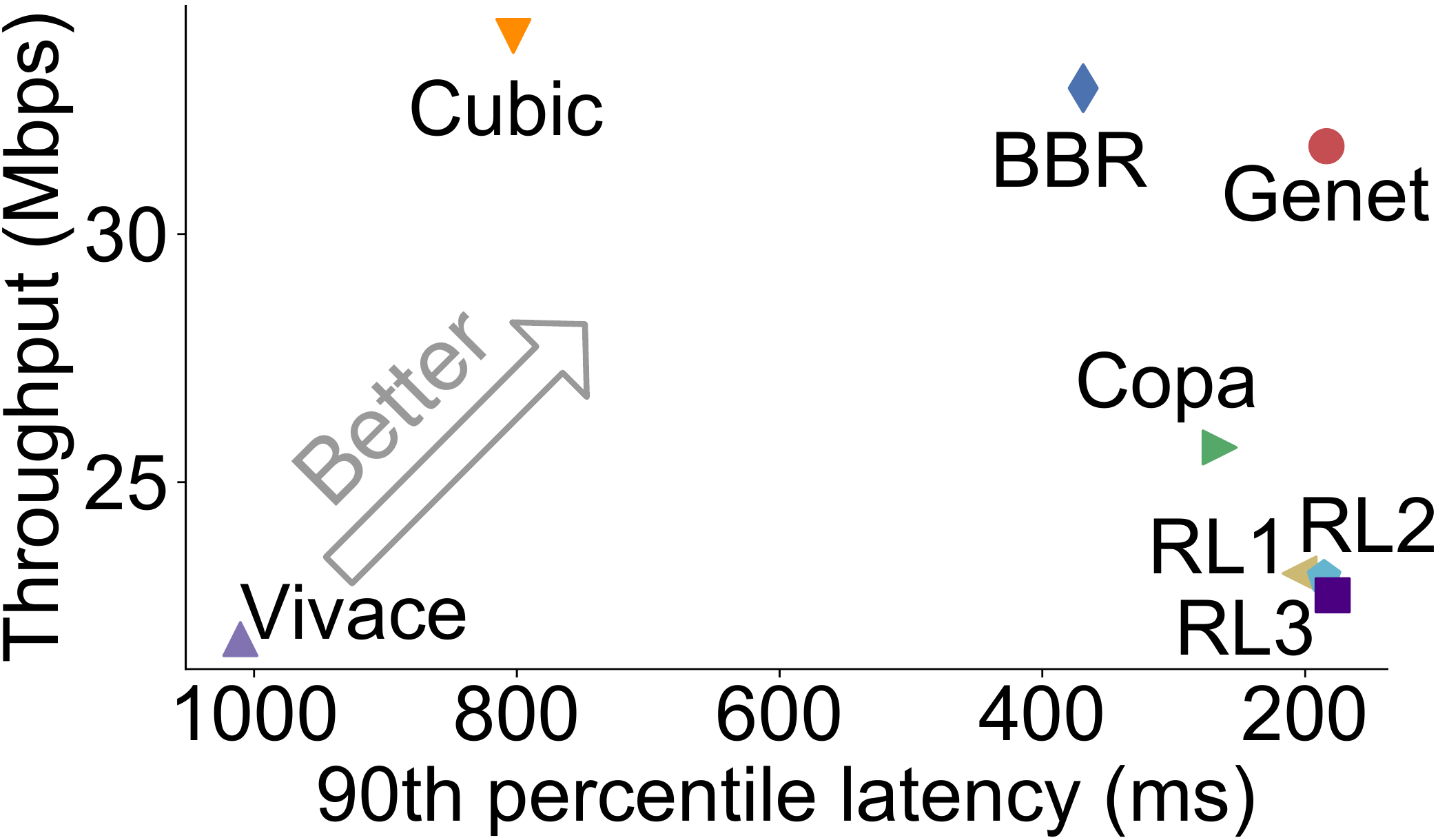}
        \caption{CC (Ethernet traces)}
        \label{fig:abr:ethernet}
    }
    \end{subfigure}
    
    \begin{subfigure}[t]{0.48\linewidth}
    {
        \includegraphics[width=0.95\linewidth]{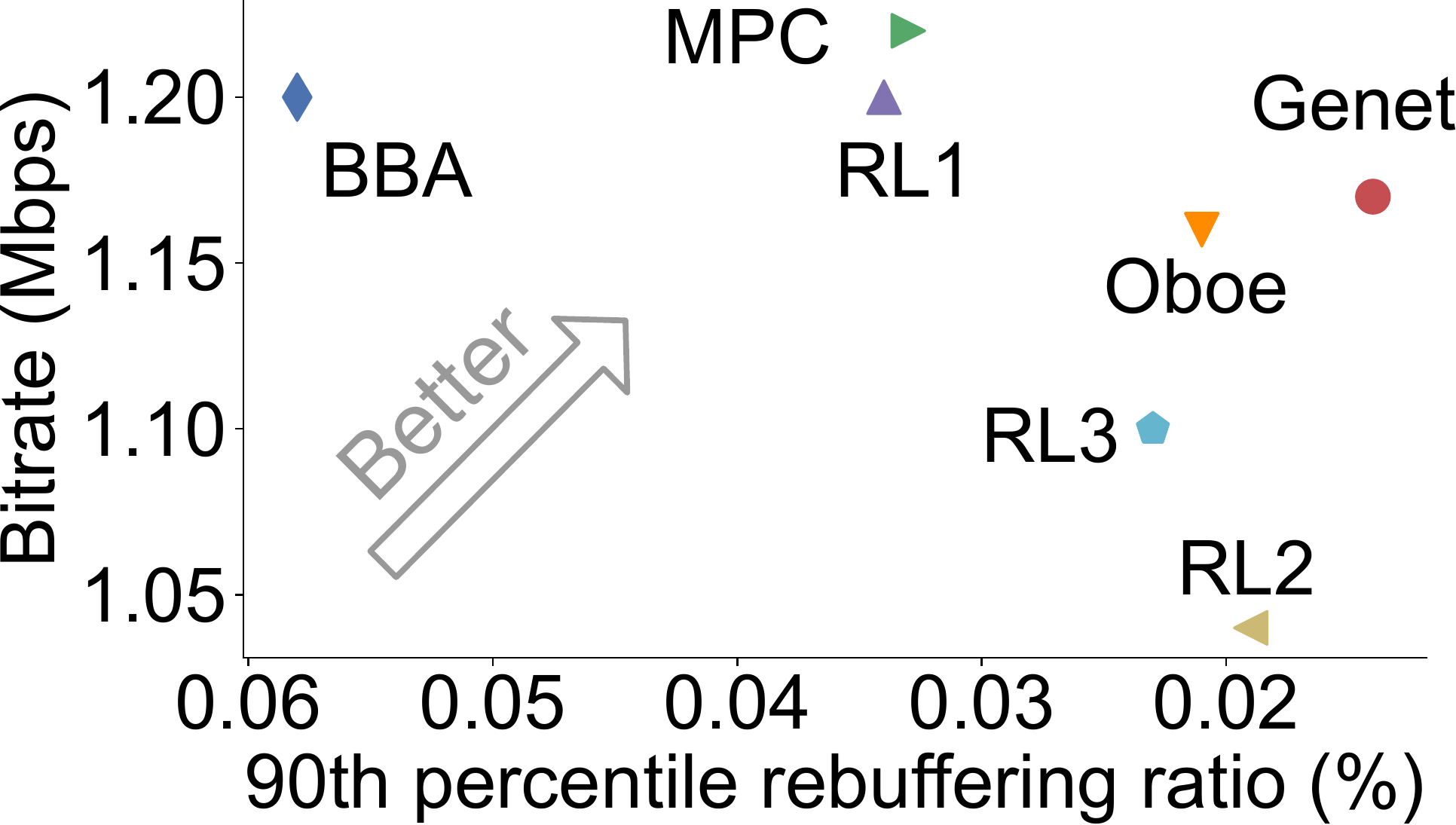}
        \caption{ABR (FCC traces)}
        \label{fig:abr:fcc}
    }
    \end{subfigure}
    \begin{subfigure}[t]{0.48\linewidth}
    {
        \includegraphics[width=0.95\linewidth]{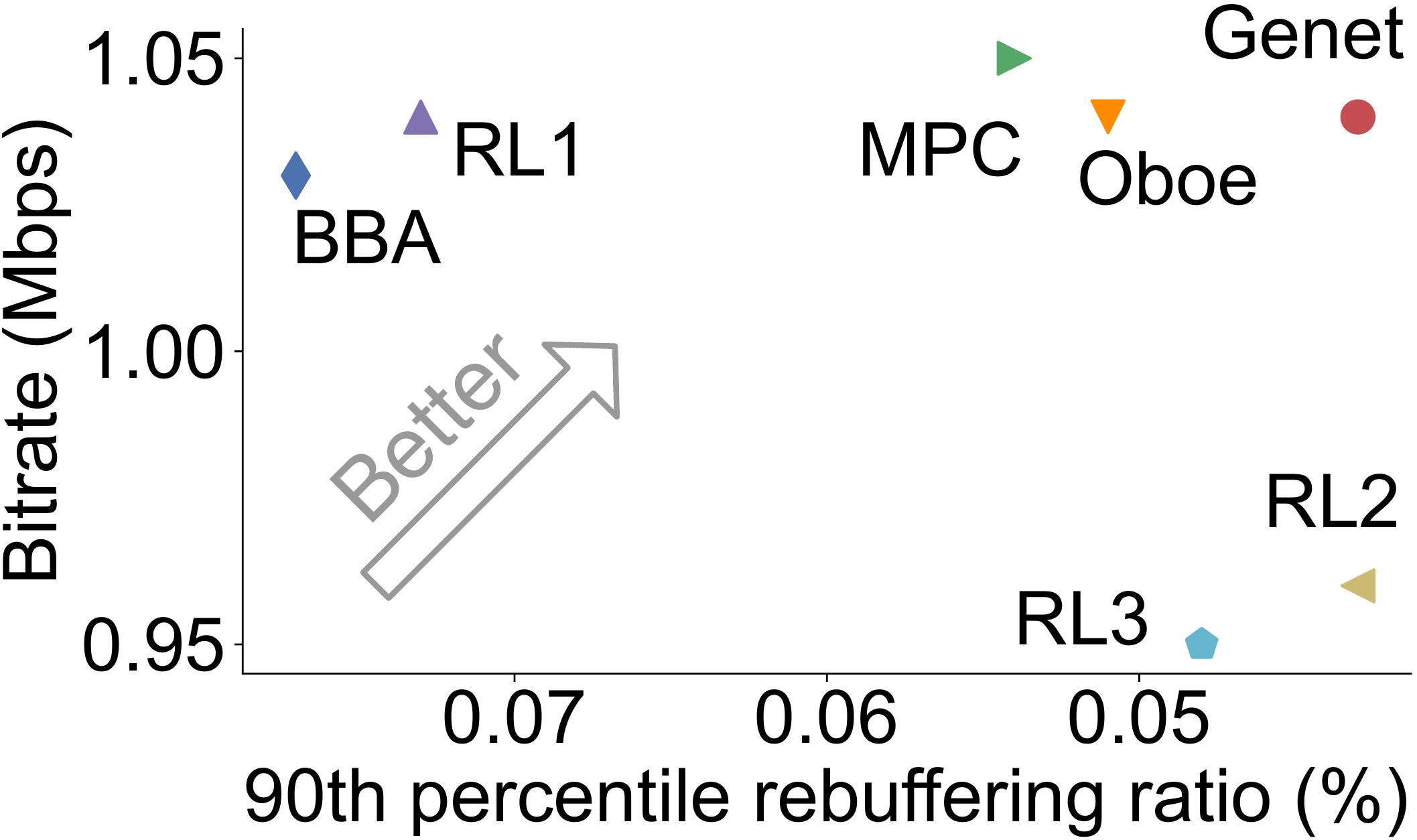}
        \caption{ABR (Norway traces)}
        \label{fig:abr:norway}
    }
    \end{subfigure}
    \tightcaption{RL-based ABR and CC vs. rule-based baselines.}
    \label{fig:two-dim}
    \vspace{3pt}
\end{figure}

\mypara{Breakdown of performance} 
Figure~\ref{fig:two-dim} takes one \name-trained ABR policy (with MPC as the rule-based baseline) and one \name-trained CC policy (with BBR as the rule-based baseline) and compares their performance with a range of rule-based baselines along with individual performance metrics.
We see that the \name-trained ABR and CC policies stay on the frontier and outperform other baselines.

\mypara{Real-world tests}
We also test the \name-trained ABR and CC policies in five real wide-area network paths  (without emulated delay/loss), between four nodes reserved from OpenNetLab~\cite{onl,opennetlab}, one laptop at home, and two cloud servers (\S\ref{app:testbed}), allowing us to observe their interactions with real network traffic.
For statistical confidence, we run the \name-trained policies and their baselines back-to-back, each at least five times, and show their performance in Figure~\ref{fig:eval:real}. \zx{The system metrics behind each reward value are shown in Table~\ref{tab:ABR-reward-breakdown} and Table~\ref{tab:CC-reward-breakdown}}.
In all but two cases, \name outperforms the baselines. 
On Path-2, \name-trained ABR has little improvement, because the bandwidth is always much higher than the highest bitrate, and the baselines will simply use the highest bitrate, leaving no room for improvement. 
On Path-3, \name-trained CC has negative improvement, because the network has a deeper queue than used in training, so RL cannot handle it well. This is an example where \name can fail when tested out of the range of training environments.
These results do not prove that the policies generalize to all environments; instead, they show \name's performance in a range of network settings.

\subsection{Understanding \name's design choices}
\label{subsec:eval:micro}
\begin{figure}[t]
    \centering
    \includegraphics[width=0.95\linewidth]{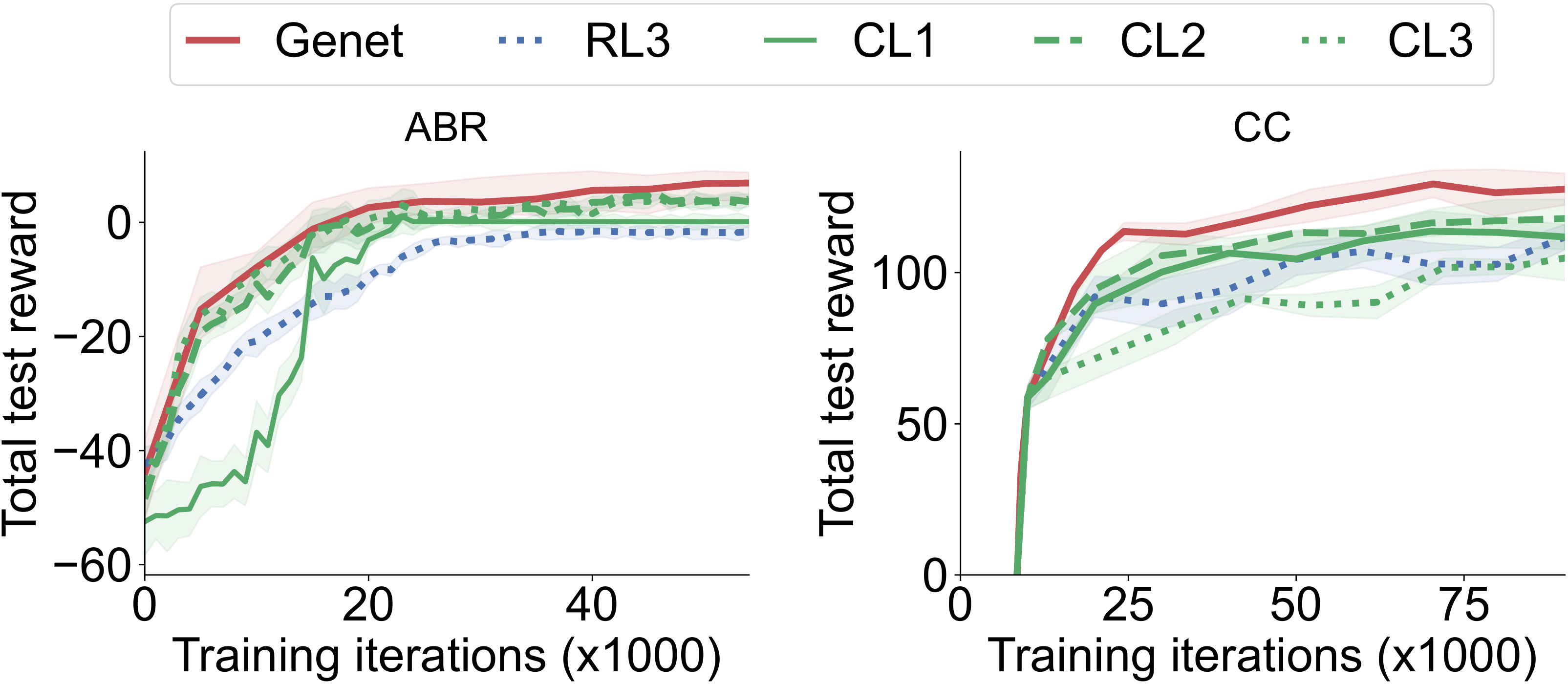}
    
    \tightcaption{\name's training ramps up faster than alternative curriculum learning strategies.}
    \vspace{4pt}
    \label{fig:eval:iterations}
\end{figure}

\mypara{Alternative curriculum-learning schemes}
Figure~\ref{fig:eval:iterations} compares \name's training curve with that of traditional RL training and three alternatives for selecting training environments described in \S\ref{sec:curricula}.
{\bf CL1} uses hand-picked heuristics (gradually increasing the bandwidth fluctuation frequency in the training environments), {\bf CL2} uses the performance of a rule-based baseline (gradually adding environments where BBR for CC and MPC for ABR performs badly), and {\bf CL3} adds traces where the current RL model \zxedit{is much worse than the optimum} 
(whereas \name picks the traces where the current RL model is much worse than a rule-based baseline).
Compared to these baselines, 
In Figure~\ref{fig:eval:iterations}, we show that \name's training curves have faster ramp-ups, suggesting that with the same number of training iterations, \name can arrive at a much better policy, which corroborates the reasoning in \S\ref{sec:curricula}.

\begin{figure}[t]
    \centering
    \includegraphics[width=0.7\linewidth]{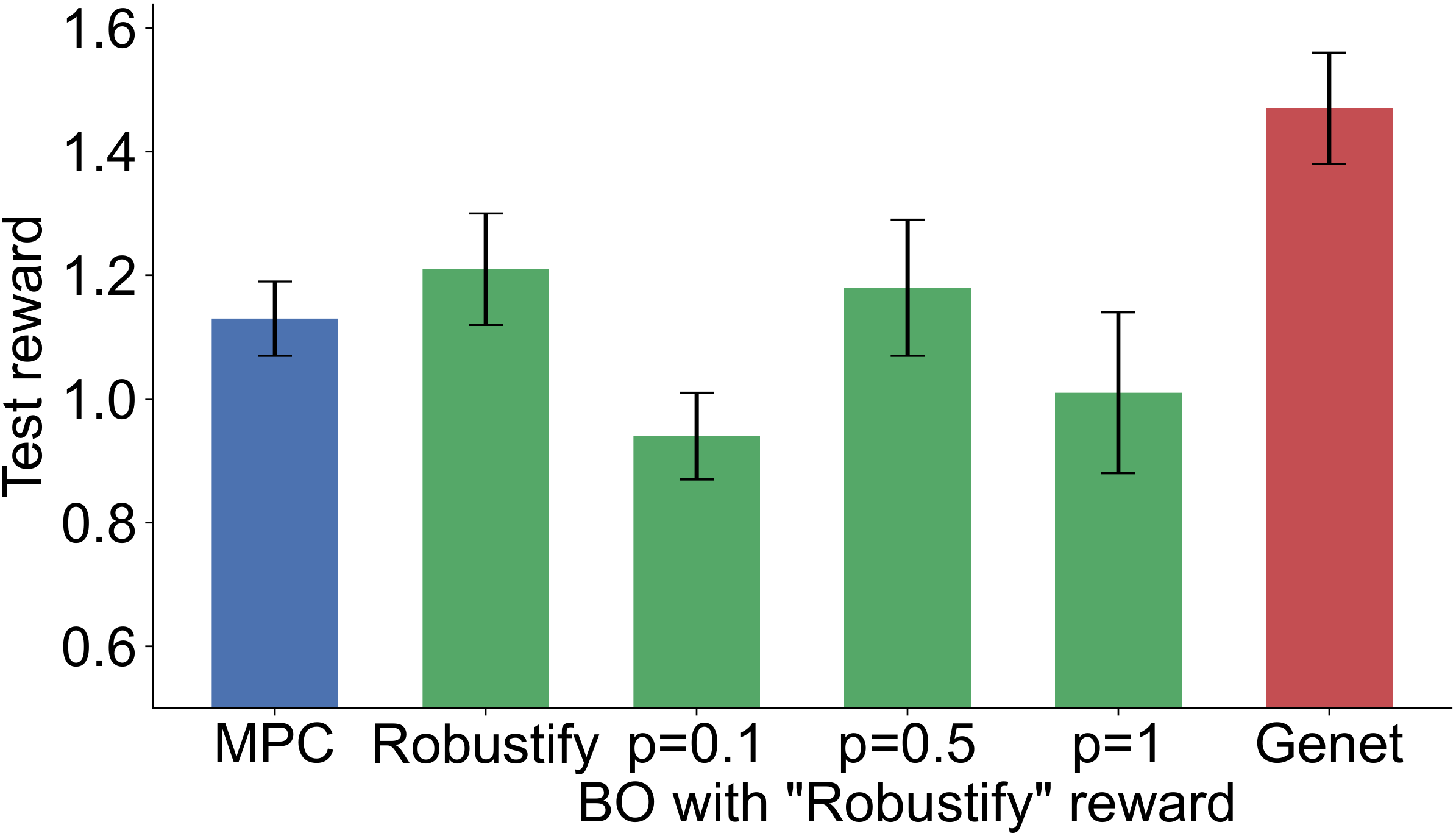}
    \vspace{2pt}
    \tightcaption{\name outperforms Robustifying~\cite{robustifying} that improves RL performance by generating adversarial bandwidth traces, and variants of \name using Robustifying's criteria in BO-based environment selection.}
    \vspace{-1pt}
    \label{fig:eval:robustifying}
\end{figure}

In addition, ``Robustifying''~\cite{robustifying}\footnote{In lack of a public implementation, we follow the description in~\cite{robustifying} (\eg non-smoothness weight) and apply it to Pensieve (with the only difference being that for fair comparisons with other baselines, we apply it on Pensieve trained on our synthetic training environments). We have verified that our implementation of Robustifying achieves similar improvements in the setting of original paper. More details are in Appendix~\ref{app:robustifying}.} (which learns an adversarial bandwidth generator)
also tries to improve ABR logic by adding more challenging environments to training. 
For a more direct comparison with \name, we implement a variant of \name where BO picks configurations that maximize the gap between RL and the optimal reward (penalized by bandwidth non-smoothness with different weights of $p$).
Figure~\ref{fig:eval:robustifying} compares the resulting RL policies with \name-trained RL policy and MPC as a baseline on the synthetic traces in Figure~\ref{fig:eval:synthetic-setting-abr}. 
We see that they perform worse than \name-trained ones and that by changing the BO's environment selection criteria, \name becomes less effective. 
\name outperforms Robustifying, because the non-smoothness metric used in~\cite{robustifying} may not completely capture the inherent difficulty of bandwidth traces (Figure~\ref{fig:difficulty-example} shows a concrete example).

\begin{figure}[t]
    \centering
    \begin{subfigure}[t]{0.48\linewidth}
    {
        \includegraphics[width=1\linewidth]{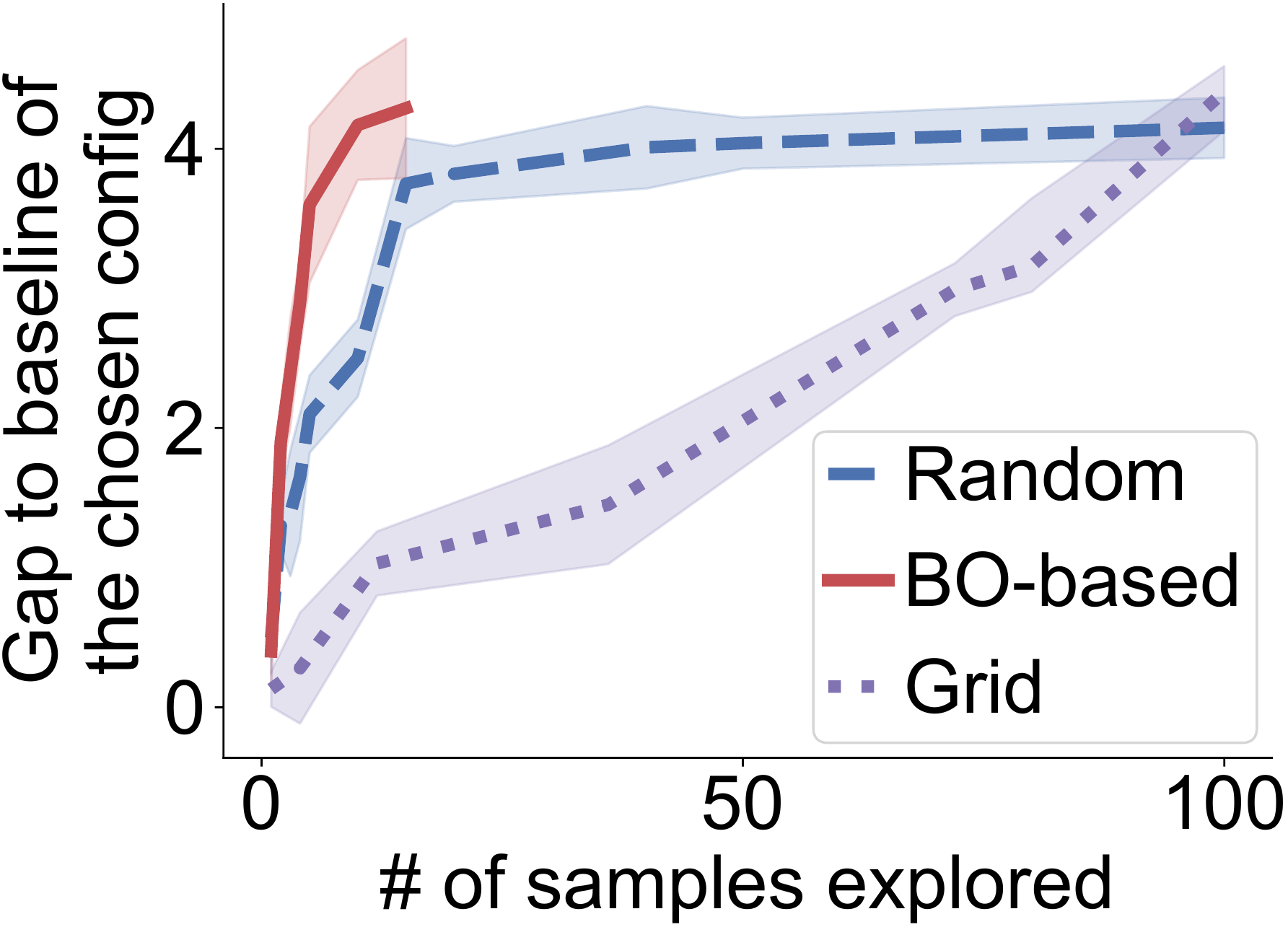}
        \caption{ABR}
        \label{}
    }
    \end{subfigure}
    \begin{subfigure}[t]{0.48\linewidth}
    {
        \includegraphics[width=1\linewidth]{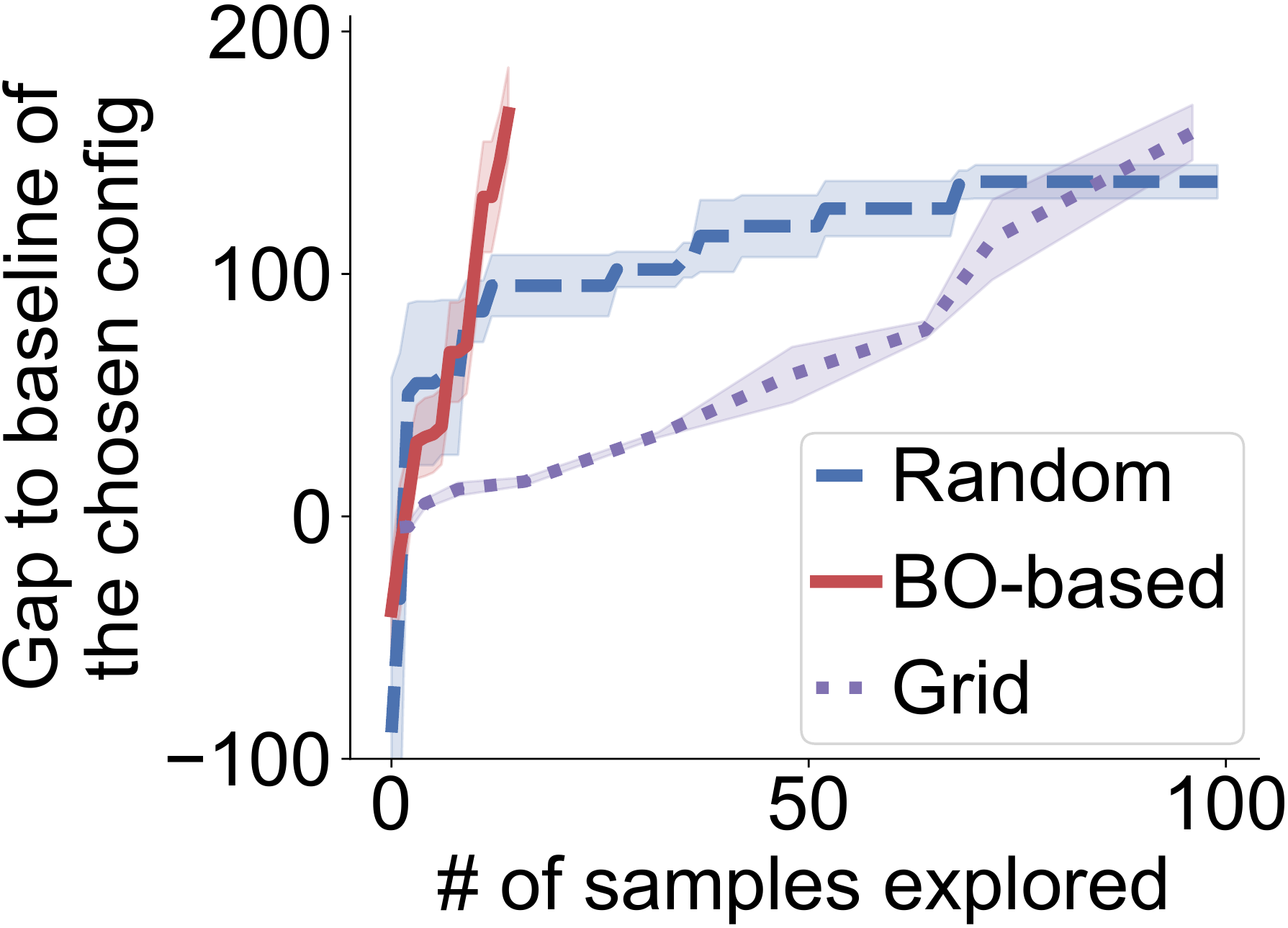}
        \caption{CC}
        \label{}
    }
    \end{subfigure}
    \vspace{2pt}
    \tightcaption{\textit{BO-based search is more efficient at finding environments with large gap-to-baselines than random exploration in the environment configuration space.} 
}
    \label{fig:eval:bo-efficiency}
\end{figure}

\mypara{BO-based search efficiency}
\name uses BO to explore the multi-dimensional environment space environment to find the environment configuration with a large gap-to-baseline. 
While BO may not always find the single optimal point in arbitrary blackbox function between environment parameters and gap-to-baseline, we found it to be a pragmatic solution.
To show it, we randomly choose an intermediate RL model during the \name training of ABR and CC.
Figure~\ref{fig:eval:bo-efficiency} shows the gap-to-baseline of the configuration selected by BO for each model within 15 search steps.
Within a small number of steps, it can identify a configuration that is almost as good as randomly searching for 100 points, which is much more expensive.
\zx{Figure~\ref{fig:eval:bo-efficiency} also includes the grid search as a reference, which starts with all configurations initialized to their respective midpoints and then searches and updates the best value for each configuration one by one. We observe that it does not converge as fast as BO.}

\section{Related work}
\label{sec:related}

\vspace{-2pt}
\mypara{Improving RL for networking}
Some of our findings regarding the lack of generalization corroborate those in previous work~\cite{remy,pensieve,aurora,robustifying,rotman2020online,dethise2019cracking}.
To improve RL for networking use cases, prior work has attempted to apply and customize techniques from the ML literature. 
For instance, \cite{robustifying} applies adversarial learning by generating relatively smooth bandwidth traces that maximize the RL regret w.r.t. optimal outcomes, \cite{verifying,kazak2019verifying} show that the generalization of RL can be improved by incorporating training environments where a given RL policy violates pre-defined safety conditions, \cite{wield,schaarschmidt2020end} incorporate randomization in the evaluation of RL-based systems, and Fugu~\cite{puffer} achieves a similar goal through learning a transmission time predictor {\em in situ}.
Other proposals seek to safely deploy a given RL policy in new environments~\cite{trainingwheel,rotman2020online,shi2021adapting}.
In many ways, \name follows this line of work, but it is different in that it systematically introduces curriculum learning, which has underpinned many recent enhancements of RL and demonstrates its benefits across multiple applications.

\mypara{Curriculum learning for RL}
There is a substantial literature on improving deep RL with curricula (\cite{narvekar2020curriculum,hacohen2019power,portelas2020automatic} give more comprehensive surveys on this subject). 
Each component of curriculum learning has been extensively studied, including how to generate tasks (environments) with potentially various difficulties~\cite{silva2018object,schmidhuber2013powerplay}, how to sequence tasks~\cite{ren2018self,sukhbaatar2017intrinsic}, and how to add a new task to training (transfer learning). 
In this work, we focus on sequencing tasks to facilitate RL training. 
It is noticed that, for general tasks that do not have a clear definition of difficulty (like networking tasks), optimal task sequencing is still an open question.
Some approaches, such as self-paced learning~\cite{kumar2010self} advocate the use of easier training examples first, while the other approaches prefer to use  harder examples first~\cite{chang2017active}.
Recent work tries to bridge the gap by suggesting that an ideal next training task should be difficult for the current model's hypothesis, while it is also beneficial to prefer easier points with respect to the target hypothesis~\cite{hacohen2019power}. In other words, we should prefer an easy environment that the current RL model cannot handle well, which confirms the intuition elaborated in Bengio's seminal paper~\cite{bengio2009curriculum}, which hypothesizes that ``it would be beneficial to make learning focus on `interesting' examples that are neither too hard nor too easy.''
\name is an instantiation of this idea in the context of networking adaptation, and the way to identify the rewarding (or ``interesting'') environments is by using the domain-specific rule-based schemes to identify where the current RL policy has a large room for improvement.

Automatic generation of curricula also benefits generalization, particularly when used together with domain randomization~\cite{peng18}.
Several schemes boost RL's training efficiency by iteratively creating a curriculum of challenging training environments (\eg~\cite{paired,adr}) where the RL performance is much worse than the optimal outcome (\ie maximal regret).
When the optimal policy is unavailable, they learn a competitive baseline~\cite{paired} to approximate the optimal policy or a metric~\cite{adr} to approximate the regret. 
\name falls in this category, but proposes a domain-specific way of identifying rewarding environments using rule-based algorithms.

Some proposals in safe policy improvement (SPI) for RL also use rule-based schemes~\cite{ghavamzadeh2016safe,laroche2019safe}, though for different purposes than \name.
While \name uses the performance of rule-based schemes to identify where the RL policy can be maximally improved, SPI uses the decisions of rule-based algorithms to avoid violation of failures during training.

\vspace{5pt}
\tightsection{Discussion}
\label{sec:discuss}

\zx{
\noindent {\bf Does a small gap-to-baseline always mean that an RL model has small improvement when trained on it?} \\
Although a small gap-to-baseline on a network environment indicates that the RL model already performs quite closely with the rule-based baseline, there is still a chance that the RL model could be greatly improved when trained in that environment.
This is because if the rule-based baseline performs very badly in an environment, the gap-to-baseline will no longer be indicative of the potential improvement of RL training.
For example, Cubic may perform poorly on a high-bandwidth link with occasional random packet loss, as Cubic does not differentiate random packet loss and congestion-induced loss, causing it to lower congestion window size when the available bandwidth does not drop.
In such cases, even if an RL model has a small gap-to-baseline with Cubic, there {\em could} still be room for the RL model to improve performance, but \name may not choose to prioritize such environments.
That said, this problem could be mitigated by using a more performant baseline or an ``ensemble'' of existing baselines (\ie measuring the maximum gap to {\em any} baseline from a set).}

\vspace{0.2cm}
\zx{
\noindent {\bf Does training in environments of large gap-to-baseline always lead to large RL model improvement?} \\
Unfortunately, the answer is not always.
RL models may not always be able to approximate the performance of rule-based baselines, \eg due to an RL model's coarse decision granularity.
For instance, Aurora (an RL-based CC) is a monitor-interval-based CC
algorithm. Each monitor interval needs to be long enough to accumulate enough
packet acks (\eg 10--50) to compute the features (throughput, latency, etc.) for the RL model to select the sending rate.
In contrast, traditional TCP algorithms like Cubic and BBR can update sending rate (cwnd) on the arrival of each packet ack.
Thus, Aurora has a much coarser decision granularity than traditional TCPs, rendering it hard for the RL model to approximate the traditional TCP's behavior when the network condition suddenly changes.
For instance, during sudden bandwidth drops and rapid queue buildups, the inter-packet interval dramatically increases, and so does Aurora's monitor interval, whereas TCP Cubic or BBR can still update its sending rate on each packet ack.
In these cases, Aurora will never ramp up or reduce sending rate as fast as its rule-based baselines, so even with a large gap-to-baseline in such environments, Aurora may not see a large reward improvement.
}

\vspace{0.2cm}
\zx{
\noindent {\bf What if a rule-based baseline does not exist?} \\
The current \name training framework requires the existence of a rule-based baseline for the target networking problem. If the problem does not have a well-studied rule-based baseline, there are three alternative training methods that \name can fall back to.
First, \name can fall back on traditional RL training. Although it loses the benefits of curriculum learning, it may still produce a reasonable RL-based policy.
Second, we can use the performance gap between an optimal solution based on ground truth knowledge (such as future bandwidth variation) and the current RL model as the guidance of rewarding network environment selection.
\cite{robustifying} trains an ABR RL model using network traces from a bandwidth-generating model. The training of the bandwidth-generating model is then guided by the performance gap between the optimal solution and the current RL model. This training method works well when the optimal solution is feasible and computationally cheap.
Third, a trained RL model can be treated as a rule-based baseline. \cite{paired} trains two RL models (with identical model architecture) competitively on the environments produced by an adversarial generator. The adversarial generator is a neural network that aims to maximize the reward difference between the two RL models.
However, the training complexity increases due to the increased number of models to be trained. Even though \name can fall back on alternative training methods, how to extend it to work in applications domains that do not have an existing rule-based baseline remains to be investigated.}

\vspace{6pt}
\tightsection{Conclusion}
We present \name, a new training framework to improve the training of deep RL-based network adaptation algorithms.
For the first time, we introduce curriculum learning to the networking domain as the key to reaching better RL performance and generalization.
To make curriculum learning efficient in networking, the main challenge is how to automatically identify the ``rewarding'' environments that can maximally benefit from retraining.
\name addresses this challenge with a simple-yet-efficient idea that highly rewarding network environments are where the current RL performance falls significantly behind that of a rule-based baseline scheme.
Our evaluation on three RL use cases shows that \name improves RL policies (in both performance and generalization) in various environments and workloads.

\mypara{Ethics}
This work does not raise any ethical issues.

\vspace{6pt}
\tightsection{Acknowledgements}
We thank Microsoft Research and OpenNetLab for their generous support of
server nodes.
We thank the SIGCOMM reviewers and our shepherd, Michael Schapira,
for their invaluable feedback. This research is partly supported
by NSF (2146496, 1901466, 2131826), UChicago CERES
Center, a Google Faculty Research Award.

\bibliographystyle{plain}
\bibliography{paper}

\appendix


\vspace{10pt}
\section{Appendices}
\label{sec:appendix}
Appendices are supporting material that has not been peer-reviewed.


\begin{table*}[t]
\begin{centering}

\scalebox{0.9}{
\begin{tabular}{lrrrrr}
  \toprule
  ABR Parameter & RL1 & RL2 & RL3 & Default & Original \\
  \midrule
  \rowcolor{Gray}
  Max playback buffer (s) & [2, 10] & [2, 50] & [2, 100] & 60 & 60 \\
  \rowcolor{Gray}
  Video chunk length (s) & [1, 4] & [1, 6] & [1, 10] & 4 & 4\\
  \rowcolor{Gray}
  Min link RTT (ms)  & [20, 30] & [20, 220] & [20, 1000] & 80 & 80\\
  \rowcolor{Gray}
  Video length (s) & [40, 45] & [40, 200] & [40, 400] & 196 & 196\\
  Bandwidth change interval (s) & [2, 2] & [2, 20] & [2, 100] & 5 & \\
  Max link bandwidth (Mbps)  & [2, 5] & [2, 100] & [2, 1000] & 5 & \\
  \bottomrule
\end{tabular}
}
\vspace{1ex}
\vspace{3ex}
\tightcaption{Parameters in ABR simulation. \disclaimer The synthetic trace generator is described in \S\ref{app:trace}.}
\label{tab:ABR-params}

\end{centering}
\end{table*}

\begin{table*}[t]
\begin{centering}

\scalebox{0.9}{
\begin{tabular}{lrrrrr}
  \toprule
  CC Parameter & RL1 & RL2 & RL3 & Default &  Original\\
  \midrule
  \rowcolor{Gray}
  Maximum link bandwidth (Mbps) & [0.5, 7] & [0.4, 14] & [0.1, 100] & 3.16 & [1.2, 6]\\
  \rowcolor{Gray}
  Minimum link RTT (ms) & [205, 250] & [156, 288] & [10, 400] & 100 & [100, 500]\\
  Bandwidth change interval (s) & [11, 13] & [8, 3] & [0, 30] & 7.5 & \\
  \rowcolor{Gray}
  Random loss rate & [0.01, 0.014] & [0.007, 0.02] & [0, 0.05] & 0 & [0, 0.05]\\
  \rowcolor{Gray}
  Queue (packets) & [2, 6] & [2, 11] & [2, 200] & 10 & [2, 2981]\\
  \bottomrule
\end{tabular}
}
\vspace{1ex}
\vspace{3ex}
\tightcaption{Parameters in CC simulation. \disclaimer The synthetic trace generator is described in \S\ref{app:trace}. The range of RL1 is defined as 1/9 of the range of RL3 and the range of RL2 is defined as 1/3 of RL3. The CC parameters shown here for RL1 and RL2 are example sets.}
\label{tab:CC-params}
\vspace{1ex}

\end{centering}
\end{table*}

\begin{table*}[t]
\begin{centering}

\scalebox{0.9}{
\begin{tabular}{lrrrrrr}
  \toprule
  LB Parameter & RL1 & RL2 & RL3 & Default & Original \\
  \midrule
  \rowcolor{Gray}
  Service rate  & [0.1, 2] & [0.1, 5] & [0.1, 10] & [0.5, 1.0, 2.0] & [2, 4]\\
  \rowcolor{Gray}
  Job size (byte)  & [100, 200] & [100, $10^3$] & [1, $10^4$] & $2000$ & [100, 1000]\\
  \rowcolor{Gray}
  Job interval (ms)  & [0.01, 0.05] & [0.01, 0.1] & [0.1, 1] & 0.1 & 0.2\\
  \rowcolor{Gray}
  Number of jobs  & [10, 100] & [10, 1000] & [10, 5000] & 2000 & 1000\\
  Queue shuffled probability  & [0.1, 0.2] & [0.1, 0.5] & [0.1, 1] & 0.5\\
  \bottomrule
\end{tabular}
}
\vspace{1ex}
\vspace{3ex}
\tightcaption{Parameters in LB simulation. \disclaimer The synthetic trace generator is described in \S\ref{app:trace}.}
\label{tab:LB-params}

\end{centering}
\end{table*}

\subsection{Details of RL implementation}
\label{app:udr}

The input of RL algorithm consists of a space of configurations, an initial policy parameters and predefined total number of iterations to train. The space of configurations is constructed by ranges of environment configurations. Each range is marked by the configuration's min and max values. Within a training iteration, each dimension of the space of configurations is uniformly sampled to create $K$ configurations. For each configuration, $N$ random environments are created. Thus, rollouts are collected by running the policy on total $K\times N$ environments to update the policy. When the policy is updated for the predefined number of iterations, the RL algorithm stops training and outputs a trained policy.
{
 \begin{algorithm}[h!]
 \small
     \caption{Traditional Reinforecment Learning (RL)}
     \label{alg:udr}
     \begin{algorithmic}[1]
     \Require
     $\Omega$: space of configurations,
     $\theta$: initial policy parameters,
     $N_{iters}$: \# of iterations
     \Ensure $\theta$: returned policy parameters
     \For{$i$ from 1 to $N_{iters}$}
         \State $\Phi_{rand}\leftarrow \emptyset$
         \For{1 \textbf{to} $K$} \Comment{$K$: \# configs per iteration}
             \State $p_i \sim Random(\Omega)$ \Comment{Uniformly sampled config in $\Omega$}
             \For{1 \textbf{to} $N$} \Comment{$N$: \# random envs per config}
                 \State $E \leftarrow S(p_i)$ \Comment{Create a simulated env by $p_i$}
                 \State \textbf{rollout} $\phi \sim \pi_\theta(\cdot ; E)$ \Comment{Rollout policy $\pi_\theta$ on $E$}
                 \State $\Phi_{rand} \leftarrow \Phi_{rand} \cup \phi$
             \EndFor
         \EndFor
         \State \textbf{with} $\Phi_{rand}$ update:
         \State   $\theta \leftarrow \theta + \nu \bigtriangledown_\theta J(\pi_\theta)$  \Comment{Gradient update with rate $\nu$}
     \EndFor
     \State\Return $\theta$
     \end{algorithmic}
 \end{algorithm}
}

\begin{algorithm}[h]
    \small
        \caption{\name training framework}
        \label{alg:gdr}
        \begin{algorithmic}[1]
        \Require 
        $\Omega$: uniform configuration distribution (equal probability on each configuration),
        $\pi^{rule}$: rule-based policy.
        \Ensure
        $\theta$: final RL policy parameters
	\Function{\name}{$\Omega,\pi^{rule}$}
        \State $\theta\leftarrow$ Random initial policy parameters
        \State $\Omega_{cur}\leftarrow\Omega$ \Comment{$\Omega_{cur}$ will be updated and used for training}
        \For{from 1 to $N_{iter}$} \Comment{\# of exploration iterations}
	    \State \textsc{BO.initialize}$(\Omega)$ \Comment{Initialize with full config space $\Omega$}\label{line:bo-start}
	    \For{from 1 to $N_{boTrials}$} \Comment{\# of trial configs by BO}
	        \State $p\leftarrow$ \textsc{BO.getNextChoice()}
	        \State $adv\leftarrow$ \textsc{CalcBaselineGap}$(p,\pi^{rl}_\theta,\pi^{rule})$
	        \State \textsc{BO.update}$(p,adv)$
	    \EndFor
            \State $p_{new}\leftarrow$\textsc{BO.getDecision()} \label{line:bo-end}\\
            \Comment{Weight new config $p_{new}$ by $w$ and old configs by $1-w$} 
            \State $\Omega_{cur}\leftarrow (1-w)\cdot\Omega_{cur}+w\cdot \{p_{new}\}$ \label{line:reweight}
            \State $\theta\leftarrow$\textsc{UniformDomainRand}$(\Omega_{cur},\theta,N_{iters})$ \label{line:udr}
        \EndFor
	\State\Return $\theta$
	\EndFunction
	\Function{CalcBaselineGap}{$p,\pi^{rl}_\theta,\pi^{rule}$}\label{line:advantage-function}
	\State \textbf{Initialize}: $samples \leftarrow \emptyset$
        \For{1 \textbf{to} $N_{Tests}$} \Comment{\# of reward comparisons}
            \State $E \leftarrow S(p)$ \Comment{Create a simulated env by $p_i$}
            \State \textbf{rollout} $\phi^{rl} \sim \pi^{rl}_\theta(\cdot ; E)$ \Comment{Rollout RL $\pi^{rl}$}
            \State \textbf{rollout} $\phi^{rule} \sim \pi^{rule}(\cdot ; E)$ \Comment{Rollout rule-based $\pi^{rule}$}
            \State add $Reward(\phi^{rule})-Reward(\phi^{rl})$ to $samples$
        \EndFor
        \State\Return \textsc{mean}$(samples)$
	\EndFunction
        \end{algorithmic}
    \end{algorithm}

\subsection{Trace generator logic}
\label{app:trace}

\mypara{ABR}
For the simulation in ABR, the link bandwidth trace has the format of [timestamp (s), throughput (Mbps)]. Our synthetic trace generator includes 4 parameters: minimum BW (Mbps), maximum BW (Mbps), BW changing interval (s), and trace duration (s). Each timestamp represents one second with a uniform [-0.5, 0.5] noise. Each throughput follows a uniform distribution between [min BW, max BW]. The BW changing interval controls how often throughput change over time, with uniform [1, 3] noise. Trace duration represents the total time length of the current trace. 

\mypara{CC} The trace generator in the CC simulation takes 6 inputs: maximum BW (Mbps), BW changing interval (s), link one-way latency (ms), queue size (packets), link random loss rate, delay noise (ms), and duration (s). It outputs a series of timestamps with 0.1s step length and dynamic bandwidth series. Each bandwidth value is drawn from a uniform distribution of range [1, max BW] Mbps. The BW changing interval allows bandwidth to change every certain seconds. The link one-way latency is used to simulate packet RTT. The queue size simulates a single queue in a sender-receiver network. Link random loss rate determines the chance of random packet loss in the network. Delay noise determines how large a Gaussian noise is added to a packet. The trace duration is determined by the duration input.

\mypara{LB}
We use the similar workload traces generator as the Park \cite{park-code} project, where jobs arrive according to a Poisson process, and the job sizes follow a Pareto distribution with parameters [shape, scale]. In the simulation, all servers process jobs from their queues at identical rates. \\

\subsection{Details of Figure~\ref{fig:motivation-optimal}}
\label{app:trace-adding}
Trace sets in Figure~\ref{fig:motivation-optimal} was generated by two
configurations. For trace set X, we used BW range: 0--5Mbps, BW changing
frequency: 0--2s. For trace set Y, we used BW range: 0--10Mbps, BW changing
frequency: 4--15s. As a motivation example, each trace set contains 20 traces to show the testing reward trend.

\subsection{Testbed setup}
\label{app:testbed}
\mypara{ABR}
To test our model on a client-side system, we first leverage the testbed from Pensieve \cite{pensieve-code}, which modifies dash.js (version 2.4) to support MPC, BBA, and RL-based ABR algorithms. We use the “Envivio- Dash3” video which format follows the Pensieve settings. In this emulation setup, the client video player is a Google Chrome browser (version 85) and the video server (Apache version 2.4.7) run on the same machine as the client. We use Mahimahi~\cite{netravali2015mahimahi} to emulate the network environments from our pre-recorded FCC~\cite{hongzi-fccdata}, cellular~\cite{norway-data}, Puffer~\cite{puffer} network traces, along with an 80 ms RTT, between the client and server. All above experiments are performed on UChicago servers.

\begin{figure}[t]
    \centering
    \includegraphics[width=\linewidth]{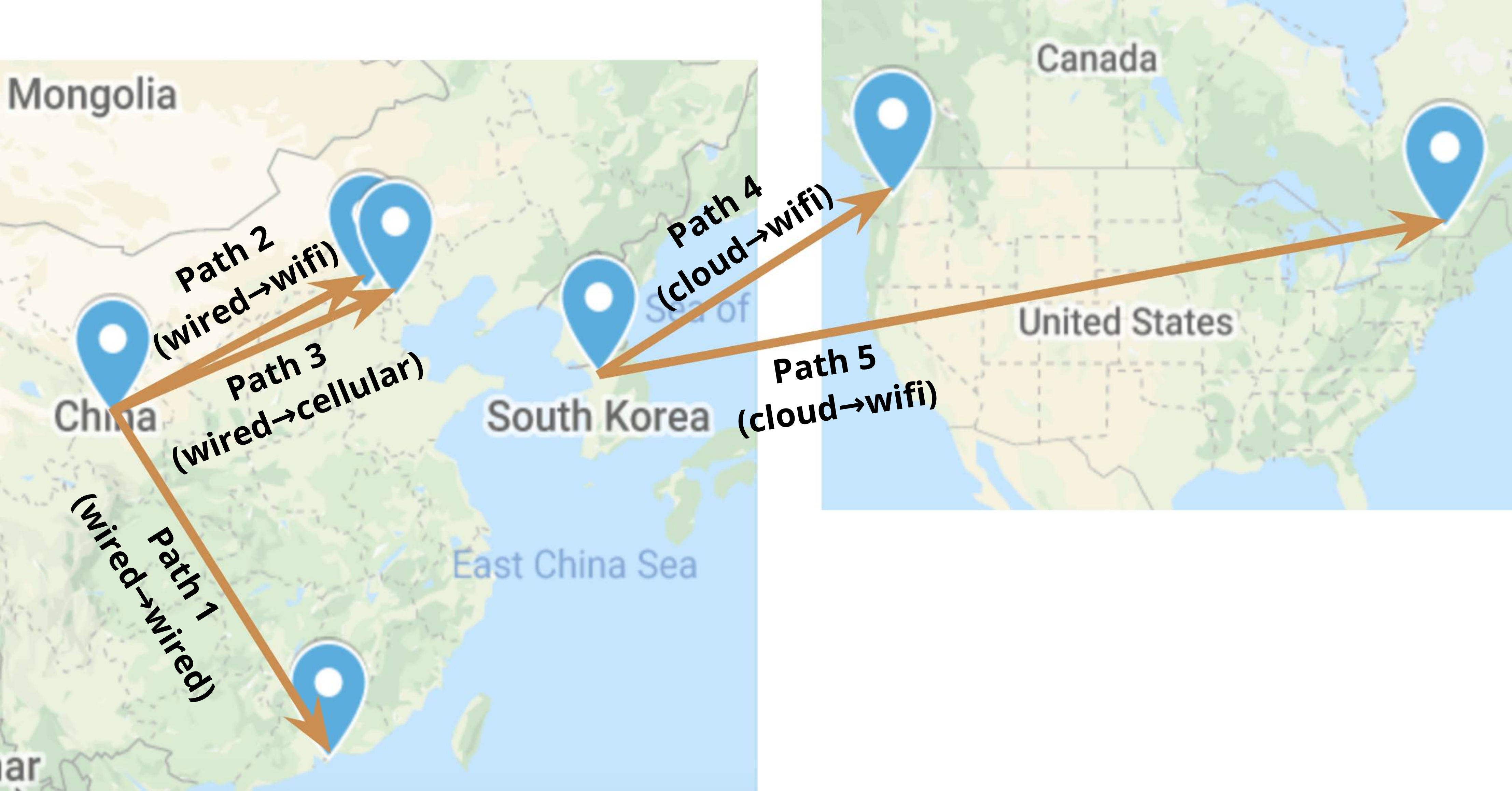}
    \vspace{-5pt}
    \tightcaption{Real-world network paths used to test ABR and CC policies.}
    \label{fig:eval:map}
\end{figure}

\mypara{CC}
We build up CC testbed on Pantheon~\cite{pantheon} platform on a Dell Inspiron 5521 machine. Pantheon uses network emulator Mahimahi~\cite{netravali2015mahimahi} and a network tunnel which records packet status inside the network link. We run local customized network emulation in Mahimahi by providing a bandwidth trace and network configurations. We run remote network experiment by deplopying pantheon platform on the nodes shown in Figure~\ref{fig:eval:map}. Among all the CC algorithms tested, BBR~\cite{bbr} and TCP Cubic~\cite{cubic} are provided by Linux kernel and are called via iperf3. PCC-Aurora~\cite{aurora} and PCC-Vivace~\cite{vivace} are implemented on top of UDP. We train our models in python and Tensorflow framework and port the models into the Aurora C++ code.

\mypara{Real network testbed}
We also test the \name-trained ABR and CC policies in real wide-area network paths (depicted in Figure~\ref{fig:eval:map}), including four nodes reserved from~\cite{onl}, one laptop at home, and two cloud servers.

\subsection{Details on reward definition}
\label{app:metrics}

\mypara{ABR} The reward function of ABR is a linear combination of bitrate, rebuffering time, and bitrate change. The bitrate is observed in kbps, and the rebuffering time is in seconds, and bitrate change is the bitrate change between bitrate of current video chunk and that of the previous video chunk. Therefore, a reward value can be computed for a video chunk. The total reward of a video is the sum of the rewards of all video chunks.

\mypara{CC} The reward function of CC is a linear combination of the throughput (packets per second), average latency (s), and packet loss (percentage) over a network connection. In training, a reward value is computed using the above metrics observed within a monitor interval. The total reward is the sum of the rewards of all monitor intervals in a connection.

\mypara{LB} The reward function of LB is the average runtime delay of a job set, which is measured by milliseconds. For each server, we observe its total work waiting time in the queue and the remaining work currently being processed. After the incoming job is assigned, the server would summarize and update the delay of all active jobs.


\begin{table*}[t]
\begin{centering}

\scalebox{0.9}{
\begin{tabular}{ccccccc}
  \toprule
  & ABR & Bitrate (Mbps) & Rebuffering (s) & Bitrate change (Mbps) & Reward \\
  \hline
   & MPC & 3.98 & 0.03 & 0.02 & 3.66\\
  Path 1 & BBA & 3.84 & 0.018 & 0.15 & 3.51\\
  & \name & 3.87 & 0.006 & 0.04 & 3.77\\
    \hline
   & MPC & 3.22 & 0.041 & 0.07 & 2.74 \\
  Path 2 & BBA  & 2.81 & 0.014 & 0.12 & 2.55\\
  & \name & 3.15 & 0.008 & 0.07 & 3.01\\
    \hline
   & MPC & 2.24 & 0.042 & 0.04 & 1.78\\
  Path 3 & BBA & 1.75  & 0.03 & 0.05 & 1.40\\
  & \name & 2.26 & 0.033 & 0.02 & 1.91\\
    \hline
   & MPC & 2.93 & 0.013 & 0.04  & 2.76\\
  Path 4 & BBA & 2.96 & 0.05 & 0.03  & 2.43\\
  & \name & 2.88 & 0.002 & 0.02 & 2.84\\
    \hline
   & MPC & 2.35 & 0.027 & 0.05  & 2.03\\
  Path 5 & BBA & 1.82 & 0.022 & 0.04 & 1.56\\
  & \name & 2.32 & 0.004 & 0.03 & 2.25\\
  \bottomrule
\end{tabular}
}
\vspace{1ex}
\vspace{1ex}
\tightcaption{Reward breakdown of Figure~\ref{fig:eval:real}(a) in ABR real-world experiment. }
\label{tab:ABR-reward-breakdown}

\end{centering}
\end{table*}

\begin{table*}[t]
\begin{centering}

\scalebox{0.9}{
\begin{tabular}{c   ccccc}
  \toprule
  Path & CC & Throughput (Mbps) & 90th percentile latency (ms) & Packet loss rate & Reward \\
  \hline
  & BBR  & 164.2 & 57.25 & 0.0906  & -35.62 \\
  Path 1 & Cubic & 158.2 & 56.60 & 0.0072 & 104.2   \\
  & \name & 180.5 & 55.54  & 0.0063 & 152.1 \\
    \hline
    & BBR & 0.2108 & 3346 & 0.0407& -1721 \\
 Path 2 & Cubic & 0.2149 & 6978 & 0.2206 & -4273 \\
  & \name & 0.1975 & 6381 & 0.0267 & -3178\\
    \hline
    & BBR & 5.40 & 1581 & 0.0136 & -705.9 \\
 Path 3 & Cubic & 6.63 & 1400 & 0.0382 & -719.1 \\
  & \name & 4.91 & 1180  & 0.0075 & -439.9 \\
  \bottomrule
\end{tabular}
}
\vspace{1ex}
\vspace{1ex}
\tightcaption{Reward breakdown of Figure~\ref{fig:eval:real}(b) in CC real-world experiments.}
\label{tab:CC-reward-breakdown}
\vspace{1ex}

\end{centering}
\end{table*}

\subsection{Baseline implementation}
\label{app:robustifying}
According to the paper~\cite{robustifying}, we train an additional RL model for Robustify to improve the main RL-policy model by generating adversarial network traces inside ABR. The state of the adversary model contains the bitrate chosen by the protocol for the previous chunk, the client buffer occupancy, the possible sizes of the next chunk, the number of remaining chunks, and the throughput and download time for the last downloaded video chunk. The action is to generate the next bandwidth in the networking trace, in order to optimize the gap between the ABR optimal policy, RL-policy, and the unsmoothness, which is the absolute difference between the last two chosen bandwidths. Here, the penalty of unsmoothness is set as 1, same as the paper.

We use PPO as the training algorithm, and train the Robustify adversary model with a RL model until they both converge. Afterward, we add the traces Robustify model generated into the RL training process to retrain the RL. The PPO parameter settings follow the original paper. 

As an alternative implementation, we also use the reward defined in Robustify as the training signal for BO to search and update environments. For the unsmoothness penalty here, we empirically tried three numbers: 0.1, 0.5, 1. From our results, penalty=0.5 works better than others.

\subsection{Reward value breakdown}
\zx{Table~\ref{tab:ABR-reward-breakdown} and Table~\ref{tab:CC-reward-breakdown} contain the system metrics behind the reward values in Figure~\ref{fig:eval:real} for ABR and CC, respectively. The breakdown is done by decomposing the reward equations introduced in Table~\ref{tab:examples}. For ABR, Table~\ref{tab:ABR-reward-breakdown} shows that \name tends to train a model that leads to less rebuffering and more smoothed bitrate selection without significantly sacrificing the average bitrate.
For CC, Table~\ref{tab:CC-reward-breakdown} shows that \name-trained model tends to have a lower 90th percentile latency and packet loss rate while not reducing throughput too much on Path 2 and 3. On Path 1, the performance gain is mainly from the larger throughput that \name-trained model enables.}

\subsection{Train RLs and CLs with more iterations}
\begin{figure}[t]
    \centering
    \includegraphics[width=0.95\linewidth]{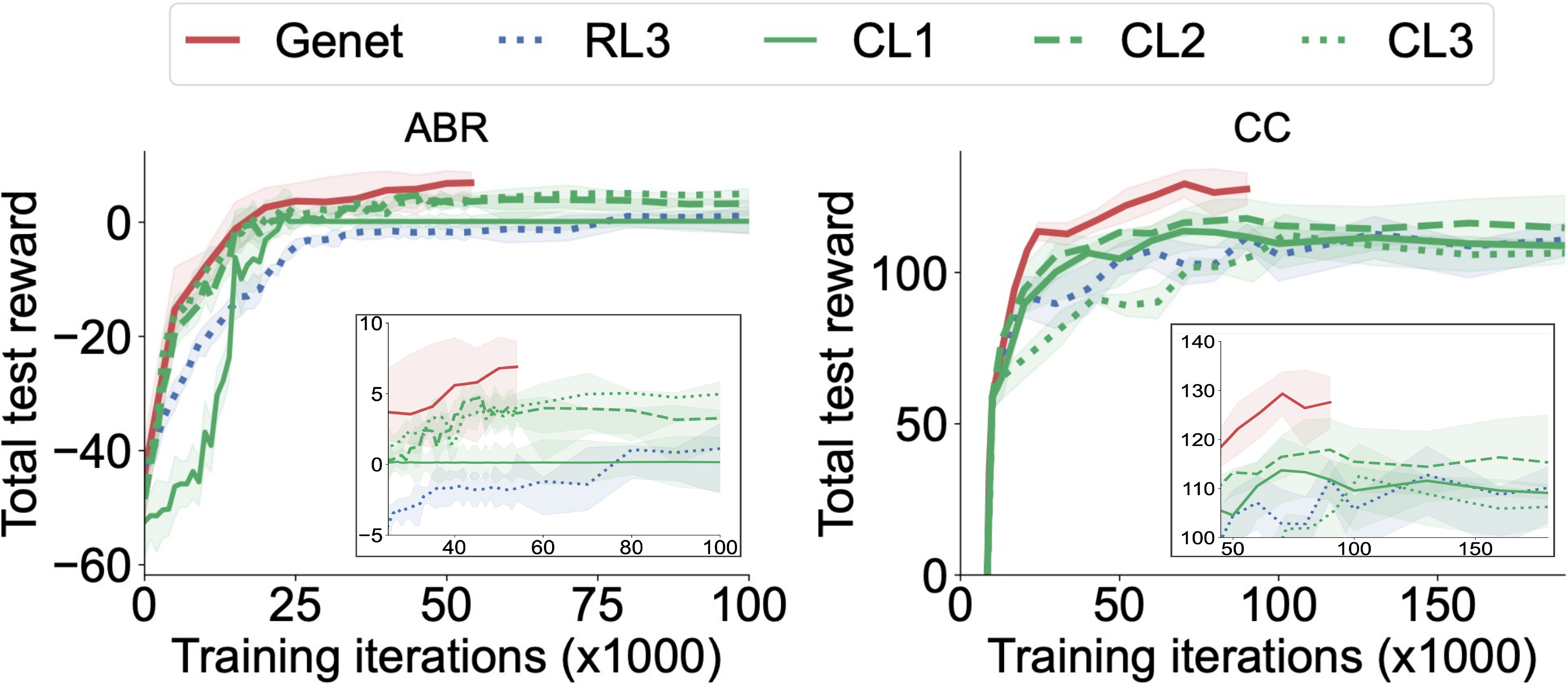}
    \vspace{0.1cm}
    \tightcaption{Training RL and CL with more iterations still cannot outperform \name.}
    \vspace{0.25cm}
    \label{fig:eval:long-run}
\end{figure}

\zx{To understand whether baselines like RLs and CLs can outperform \name if they are given more training iterations, we trained RLs and CLs with twice as many training iterations as \name. We empirically found that training with more iterations did not help the models trained by RLs and CLs as much as those trained by \name. Their learning curves are shown in Figure~\ref{fig:eval:long-run}.}

\end{document}